\documentclass[article,a4paper]{IEEEtran}

\IEEEoverridecommandlockouts

\usepackage{graphicx}
\usepackage{amssymb, amsmath, amsfonts, amsthm}
\usepackage{multirow}
\usepackage{bm}
\usepackage{nomencl}
\usepackage{mcite}
\usepackage[noadjust]{cite}
\usepackage{setspace}
\usepackage{flushend}
\usepackage{xcolor}
\usepackage{acronym}
\usepackage{nomencl}
\usepackage{etoolbox}

\renewcommand\nomgroup[1]{%
  \item[\bfseries
  \ifstrequal{#1}{A}{General Variables}{%
  \ifstrequal{#1}{B}{Current Controller}{%
  \ifstrequal{#1}{C}{Grid-Following}{%
  \ifstrequal{#1}{D}{Grid-Forming}{%
  }}}}%
]}

\acrodef{rocof}[RoCoF]{Rate of Change of Frequency}
\acrodef{pll}[PLL]{Phase-Locked Loop}
\acrodef{vff}[VFF]{Voltage Feed-Forward}
\acrodef{vsm}[VSM]{Virtual Synchronous Machine}
\acrodef{pfr}[PFR]{Primary Frequency Regulator}
\acrodef{cf}[CF]{Complex Frequency}
\acrodef{gfm}[GFM]{Grid-Forming}
\acrodef{gfl}[GFL]{Grid-Following}

\usepackage{xparse,xcolor}

\ExplSyntaxOn

\NewDocumentCommand{\setupbibcolors}{m}
 {
  \cs_set_protected:Npn \bibitem ##1
   {
    \color{ \str_case:nnF { ##1 } { #1 } { black } }
    \heba_bibitem:n { ##1 }
   }
 }

\cs_set_eq:NN \heba_bibitem:n \bibitem

\ExplSyntaxOff

\newcommand{\rf}[1]{#1_{\rm ref}}
\newcommand{\cp}[1]{\overline{#1}}
\newcommand{\cpc}[1]{\overline{#1}^{*}}
\newcommand{\cpd}[1]{\dot{\overline{#1}}}
\newcommand{\fq}[1]{\overline{\eta}_{#1}}
\newcommand{\fqc}[1]{\overline{\eta}^{*}_{#1}}
\newcommand{\cpdc}[1]{\dot{\overline{#1}}^{*}}
\newcommand{\ii}{\imath}
\newcommand{\jj}{\jmath}
\newcommand{\ejd}{{\rm exp}(\jj \, \delta)}
\newcommand{\ejdc}{{\rm exp}(-\jj \, \delta)}

\newcommand{\jdd}{\jj \, \dot{\delta}}
\newcommand{\kpi}{\kappa_{\scriptscriptstyle \rm PI}}
\newcommand{\pll}{{\scriptscriptstyle \rm PLL}}
\newcommand{\vsm}{{\scriptscriptstyle \rm VSM}}
\newcommand{\coi}{{\scriptscriptstyle \rm COI}}

\makenomenclature

\begin{document}

\title{Taxonomy of Power Converter Control Schemes based on the
  Complex Frequency Concept}
\author{Dionysios~Moutevelis, Javier Rold\'{a}n-P\'{e}rez, {\em
    Member, IEEE} \\ Milan Prodanovic {\em Member, IEEE}, and
  Federico~Milano,~\IEEEmembership{Fellow,~IEEE}%
  \thanks{D.~Moutevelis, J. Rold\'{a}n-P\'{e}rez, and M. Prodanovic
    are with the Electrical Systems Unit, IMDEA Energy,
    Avda. Ram\'{o}n de la Sagra 3, 28935, M\'{o}stoles, Madrid, Spain.
    E-mail: dionysios.moutevelis@imdea.org, javier.roldan@imdea.org,
    milan.prodanovic@imdea.org.  D.~Moutevelis is also with Alcal\'{a}
    de Henares University, Alcal\'{a} de Henares 28801 Madrid,
    Spain.}%
  \thanks{F.~Milano is with School of Electrical and Electronic
    Engineering, University College Dublin, Belfield Campus, Dublin 4,
    D04V1W8, Ireland.  E-mail: federico.milano@ucd.ie}%
  \thanks{This work is partly supported by the Community of Madrid,
    research project PROMINT-CM (P2018/EMT4366), and Juan de la Cierva
    Incorporaci\'{o}n program (IJC2019-042342-I) from the Spanish
    Government by funding D.~Moutevelis, J. Rold\'{a}n-P\'{e}rez, and
    M. Prodanovic and by the Sustainable Energy Authority of Ireland
    (SEAI) by funding F.~Milano under project FRESLIPS, Grant
    No. RDD/00681.}%
}%

\maketitle

\IEEEpubidadjcol

\begin{abstract}
  This paper proposes a taxonomy of power converter control schemes
  based on the recently proposed concept of \textit{complex
    frequency}.  This quantity captures local frequency variations due
  to the change of both the phase angle and amplitude of bus voltages
  and current injections.  The paper derives the analytical
  expressions of the link between complex power variations and complex
  frequency of each converter controller as well as the identification
  of critical control parameters.  The main contribution of this work
  is to provide a general framework that allows classifying converters
  synchronization mechanisms and controllers.  This framework also
  allows comparing converters with synchronous machines.  To validate
  the theoretical results, extensive simulations are performed using a
  modified version of the WSCC 9-bus system.  Examples of how the
  theoretical formulations of the paper can be used to improve power
  converter control in power system applications are showcased.
\end{abstract}
 
\begin{IEEEkeywords} 
  Complex frequency, grid-following, grid-forming, power converter,
  synchronization, frequency control.
\end{IEEEkeywords}

\printnomenclature
\nomenclature[A, 01]{\(\cp{u}  \)}{Complex signal or variable}
\nomenclature[A, 02]{\(\cpc{u}  \)}{Complex conjugate of variable $\cp{u}$}
\nomenclature[A, 03]{\(\cpd{u}  \)}{Derivative of variable $\cp{u}$}
\nomenclature[A, 04]{\(\fq{u}   \)}{Complex frequency of variable $\cp{u}$}
\nomenclature[A, 05]{\(\rho_u ,  \omega_u   \)}{Real and imaginary part of complex frequency}
\nomenclature[A, 06]{\(\cp{u}'  \)}{Variable $\cp{u}$ expressed in the local reference frame}
\nomenclature[A, 07]{\(\cp{v}_h  \)}{Voltage at bus $h$}
\nomenclature[A, 08]{\(\cp{\ii}_h  \)}{Current injected at bus $h$}
\nomenclature[A, 09]{\(\cp{s}_h  \)}{Instantaneous complex power injected at bus $h$}
\nomenclature[A, 10]{\(p_h, q_h  \)}{Active and reactive power injected at bus $h$}
\nomenclature[A, 11]{\(\delta \)}{Angle for transformation between reference frames}
\nomenclature[A, 12]{\( \omega_n, v_n \)}{Nominal frequency and voltage of a bus}
\nomenclature[B, 13]{\(L_f,R_f  \)}{Converter output inductance and resistance}
\nomenclature[B, 14]{\(\cp{v}_t  \)}{Current controller output voltage}
\nomenclature[B, 15]{\(K_p, K_i  \)}{Proportional and integral constants}
\nomenclature[B, 16]{\(\cp{x}  \)}{PI controller internal state}
\nomenclature[C, 17]{\(\rf{\cp{\ii}} \)}{Current reference}
\nomenclature[C, 18]{\(K_p^{\pll} K_i^{\pll} \)}{Proportional and integral constants of PLL}
\nomenclature[C, 19]{\(\rf{\cp{s}} \)}{Complex power reference}
\nomenclature[C, 20]{\(\cp{Y}_v \)}{Virtual admittance}
\nomenclature[C, 21]{\(G_v, B_v \)}{Virtual conductance and susceptance}
\nomenclature[C, 22]{\(K_p^{\rm dc}, K_i^{\rm dc} \)}{Proportional and integral constants of outer $\rm dc$ voltage controller}
\nomenclature[C, 23]{\(K_p^{\rm ac}, K_i^{\rm ac} \)}{Proportional and integral constants of outer $\rm ac$ voltage controller}
\nomenclature[C, 24]{\(x_{\rm dc}, x_{\rm ac} \)}{Internal states of the outer $\rm dc$ and $\rm ac$ voltage controllers}
\nomenclature[D, 25]{\(\cp{x}_v
 \)}{Internal state of the voltage controller}
\nomenclature[D, 26]{\(K_p^v, K_i^v \)}{Proportional and integral constants of voltage controller}
\nomenclature[D, 27]{\(\rf{p}, \rf{q} \)}{Active and reactive power references}
\nomenclature[D, 28]{\(m_p, m_q \)}{$P/f$ and $Q/V$ droop gains}
\nomenclature[D, 29]{\(T_f \)}{Time constant of the droop controller low-pas filter}
\nomenclature[D, 30]{\(J_v,  D_p \)}{Moment of virtual inertia coefficient and virtual damping coefficient}
\nomenclature[D, 31]{\(\psi_v \)}{Virtual flux}
\nomenclature[D, 32]{\(K_Q,  D_Q \)}{Integral and droop gains of VSM voltage controller}

\section{Introduction}
\label{sec.introduction}

\subsection{Motivations}

The high penetration of converter-interfaced energy resources in
modern power grids makes necessary the contribution of these devices
to the frequency regulation of the
network~\cite{hatziargyriou2017contribution}.  A variety of such
converter-based, frequency regulating strategies has been proposed in
the literature~\cite{tayyebi2020frequency, 9408354}.  The stability
and transient operation of these controllers is well studied and
documented~\cite{wang2020grid, tayyebi2020frequency}.  However, the
contribution of each controller to the frequency at the converter
connection point has not been yet fully discussed.  In this paper, the
recently proposed concept of \ac{cf} is employed to fill precisely
this gap and discuss, through a rigorous analytical appraisal, the
effect of different power converter controllers on the frequency at
the converter connection bus.  The results are organized in a
systematic taxonomy of power converter synchronization and control
schemes.

\subsection{Literature Review}
Converter control schemes are commonly grouped into the \ac{gfl} or
\ac{gfm} categories~\cite{rocabert2012control, li2022revisiting}.  The
former are meant to measure or estimate the frequency of the grid
while passively injecting the power that is requested from them.  The
synchronization is usually achieved by means of a \ac{pll}
device~\cite{li2022revisiting, wang2020grid}.  Extensive research
regarding \ac{pll} stability and synchronization capability,
especially in weak grids, has been carried
out~\cite{wu2022synchronization, chen2022impact}.  However, \ac{gfl}
converters with \acp{pll} are commonly assumed to not affect the
frequency at their connection point.  \ac{gfm} converters do not rely
on measuring or estimating the grid voltage. Instead, they achieve
synchronization with the grid by varying the active power
injection~\cite{wang2020grid}.  Various power-based synchronization
strategies can be found in the literature~\cite{wang2020grid,
  tayyebi2020frequency, 9408354}.  Although the optimal \ac{gfm}
control method is still an open research field, the most common ones
found both in the literature and in practical applications are the
\emph{droop control} and the \emph{\ac{vsm}}~\cite{4118327,
  moran2021influence, zhong2010synchronverters, roldan2019design,
  gonzalez2021design}.  Droop controlled converters adjust their
active power output through an active power-frequency ($P/f$) droop
rule to achieve synchronization~\cite{4118327, moran2021influence}.
\ac{vsm}-based converters fully emulate a swing equation within their
control structure, typically of second
order~\cite{zhong2010synchronverters, roldan2019design,
  gonzalez2021design}.  Many \ac{vsm} implementations exist with all
recent works highlighting the necessity of inner current and voltage
controllers in a cascaded configuration~\cite{d2013virtual}.  These
cascaded loops allow the explicit inclusion of voltage and current
limitations that are necessary for the safe operation of the
converters~\cite{d2015virtual}.  Moreover, dedicated voltage control
loops are required for the operation of the converter as a voltage
source~\cite{wang2020grid}.  It is thus of interest to quantify the
effect of all these internal~(current, voltage, \ac{pll}) and
external~(droops, \ac{vsm}) controllers on the frequency at the
converter ac bus.

The precise definition of the frequency of a power system is an open
research topic which has recently received renewed
attention~\cite{kirkham2018defining}.  In~\cite{milano2016frequency},
the point was raised that frequency is not uniform in the whole
network, especially in transient conditions, and a formula to estimate
the frequency at each system bus is proposed.  In~\cite{freqcomplex},
the concept of \ac{cf} is proposed as an extension of the well-known
definition of frequency as the time derivative of the argument of a
sinusoidal signal~\cite{iec2018ieee}.  This complex variable
quantifies the change of the network frequency caused by the variation
of both the phase angle and the magnitude of the voltage.  The
interpretation of \ac{cf} is approached in the literature from various
viewpoints~\cite{paradoxes}.  The imaginary part of the frequency is,
in effect, the conventional quantity that is commonly utilized, in
signal processing and time-frequency analysis, to define the
\textit{instantaneous frequency} of a signal.  Using the same signal
processing approach, the real part of the frequency can be defined as
\textit{instantaneous bandwidth}~\cite{Cohen:1995}.  The geometric
approach, presented in \cite{freqgeom}, assumes that the voltage
(current) is \textit{the speed of a trajectory in space}.  This is
supported by the fact that the voltage (current) is the time
derivative of a time-varying flux (electric charge). This approach
leads to define the real and imaginary parts of the \ac{cf} as the
symmetrical and anti-symmetrical components, respectively, of the time
derivative of the voltage (current).  Finally, in \cite{freqfrenet},
based again on the analogy between voltage (current) and the speed of
a trajectory, the real part of the \ac{cf} is interpreted as a
\textit{radial speed}, whereas the imaginary part is interpreted as an
\textit{azimuthal speed}.

\ac{cf} has been utilized to develop novel approaches in power system
state estimation~\cite{zhong2022line} as well as to study the
synchronization stability of converters using dispatchable virtual
oscillator control~\cite{colombino2019global, he2022complex}.
However, the various converter control schemes that are found in the
literature have not been studied, thus far, under the lens of \ac{cf}.
This paper aims at filling this gap.  In \cite{dervf}, the magnitude
of the complex frequency was utilized as a metric to compare the
performance of converter primary frequency and voltage controllers.
In the same spirit, the present work is an application of the concept
of complex frequency to the characterization of the dynamic behavior
of the different parts that form the synchronization of the control of
power electronics converters.

\subsection{Contributions}

The contributions of this paper are summarized as follows:
\begin{itemize}
\item The notion of voltage \ac{cf} presented in~\cite{freqcomplex} is
  extended to other variables and signals. It is explained how \ac{cf}
  can be used as derivative operator for these signals, e.g., currents
  or voltage references.
\item Then, the generalised \ac{cf} quantity, i.e. resulting from the
  extension above, is used to derive the effect of different converter
  control schemes on the grid frequency. This novel approach allows
  the decoupling of the effect of different participating controllers
  and the identification of critical control parameters.  These
  parameters can then be tuned appropriately to maximize or minimize
  the controller effect on the grid frequency.
\item The local frequency as perceived by the converter is calculated
  and categorized.  This \textit{internal} frequency differs from the
  bus frequency due to the action of the converter controllers and
  synchronization mechanism. A parallel with the rotor speed of a
  synchronous machine is drawn, allowing the study of the two
  generation types (synchronous and asynchronous) using the same
  theoretical tools. Examples of how the internal frequency can be
  used to improve converter control are included.
\end{itemize}
The theoretical contributions of this work are supported by extended
simulations taking into account various controller configurations,
both \ac{gfl} and \ac{gfm}, and the operation of a realistic grid
benchmark.

\subsection{Paper Organization}

The organization of the rest of the paper is as follows.  In
Section~\ref{sec.prelim}, the theoretical framework for the use of
\ac{cf} as a derivative operator is established.  In
Section~\ref{sec.general}, some general mathematical derivations that
establish the relationship between frequency and complex power are
presented.  Then, they are used to define some common special cases in
power systems as well as to analyze the fundamental current controller
of power converters.  Sections~\ref{sec.gfl} and~\ref{sec.gfm} present
the effect on \ac{cf} of \ac{gfl} and \ac{gfm} control schemes,
respectively.  Section~\ref{sec.simresults} validates the theoretical
results with a case study based on the modified WSCC 9-bus system.
Lastly, Section~\ref{sec.conclusion} draws conclusions and outlines
future work.

\section{Complex Frequency as a Derivative Operator}
\label{sec.prelim}

As it is well known, a complex quantity $\cp{u}$ can be written either
in polar or rectangular coordinates, as follows:
\begin{equation}
  \label{eq:u}
  \cp{u} = u_d + \jj \, u_q = u \, {\rm exp}(\jj \varphi) \, ,
\end{equation}
where $u = \sqrt{u_d^2 + u_q^2}$ and $\varphi = \arctan(u_q/u_d)$.
For $u \neq 0$, the polar form can be also written as:
\begin{equation}
  \label{eq:u2}
  \cp{u} = {\rm exp}\big (\ln(u) + \jj \varphi \big ) \, .
\end{equation}
Throughout this paper, the annotation from \eqref{eq:u} will be used
to signify all complex quantities referred to a $\rm dq$-axis
reference frame, including currents, voltages, control signals etc.
The time derivative of \eqref{eq:u2} gives:
\begin{equation}
  \label{eq:udot}
  \cpd{u}
  = \left ({\dot{u}}/{u} + \jj \dot{\varphi} \right ) \, \cp{u}
  = (\rho_u+\jj \omega_u) \, \cp{u}
  = \fq{u} \, \cp{u} \, .
\end{equation}
The quantity $\fq{u}$ has been utilized in \cite{freqcomplex} to
define the concept of \ac{cf} of time-dependent Park vectors of the
voltage and of the injected current at a bus $h$, as follows:
\begin{equation}  
  \label{eq:vhdot}
  \begin{aligned}
    \cpd{v}_h = \fq{v} \, \cp{v}_h \, , \qquad
    \cpd{\ii}_h = \fq{\ii} \, \cp{\ii}_h \, .
  \end{aligned}
\end{equation}
The \ac{cf} includes a real part, which represents a
\textit{translation} and depends only on the magnitude of the Park
vector; and an imaginary part, which represents a \textit{rotation}
and depends only on the phase angle of the Park vector.  These two
quantities can be viewed as a special case of the symmetric and
anti-symmetric components of the more general concept of geometric
frequency defined in \cite{freqgeom}.  The advantage of defining the
real part of the \ac{cf} is twofold: (i) it provides a quantity that
has the same transient nature and same units as the conventional
instantaneous frequency, and is thus directly comparable to it; and
(ii) it allows a consistent description and formulation of the dynamic
effect of the devices that are connected to the grid.

Note that, in general and in transient conditions,
$\fq{v} \neq \fq{\ii}$, as discussed in \cite{freqcomplex}.  Moreover,
note that quantities $\fq{v}$ and $\fq{\ii}$ represent the variations
of voltage and current frequency, respectively, from the nominal,
steady-state values $\fq{vn}=\fq{\ii n}=0+\jj \omega_n$, where
$\omega_n=2 \, \pi f_n$ is the nominal frequency of the grid.

The rotation of the $dq$-axis coordinates has an important effect on
the \ac{cf} if this rotation is time dependent.  Relevantly, this is
the case of synchronous machines \cite{freqcomplex}, and converters
that are synchronized to the grid by means of a synchronization
control strategy~(active power synchronization for \ac{gfm}
converters~\cite{4118327, d2015virtual}, or \ac{pll} for \ac{gfl}
converters~\cite{li2022revisiting, wang2020grid}).  The rotation from
the local to the grid reference frame can be written as:
\begin{equation}
  \label{eq:rot}
  \begin{aligned}
    \cp{v}_h = \ejd{} \, \cp{v}_h' \, , \qquad
    \cp{\ii}_h = \ejd{} \, \cp{\ii}_h' \, ,
  \end{aligned}
\end{equation}
where $'$ denotes variables in the local (i.e., device-side) reference
frame and $\delta$ is the angle variation between the two reference
frames.  For example, in the case of the synchronous machine, $\delta$
refers to the transformation from the rotor reference frame to the
stator while for the case of converters, it transforms quantities from
the converter reference frame to the network reference frame using the
angle of the controller.  Note that the rotation \eqref{eq:rot} is
power invariant as it does not vary voltage and current magnitudes.

Differentiating the expression of $\cp{v}_h$ in \eqref{eq:rot} gives:
\begin{equation}  
  \label{eq:vdot}
  \begin{aligned}
    \cpd{v}_h' &=
    -\jdd{} \, \ejdc{} \, \cp{v}_h  +
    \ejdc{} \, \fq{v} \, \cp{v}_h \, , \\
    &= (\fq{v} - \jdd{}) \, \cp{v}_h' \, .
  \end{aligned}
\end{equation}
Hence, the \ac{cf} of a voltage referred to a local rotating reference
frame is:
\begin{equation}  
  \label{eq:vdot2}
  \fq{v}' = \fq{v} - \jdd{} \, .
\end{equation}
A similar expression can be obtained for the current:
\begin{equation}  
  \label{eq:idot2}
  \fq{\ii}' = \fq{\ii} - \jdd{} \, .
\end{equation}

In the remainder of the paper, the operators $\fq{}$ and $\fq{}'$ are
utilized to indicate the time derivative of relevant time-dependent
complex quantities (e.g., the state variables and reference currents
of the converter controllers) on grid and local reference frames,
respectively.

\section{General Derivations and Current Control}
\label{sec.general}

Using Park vectors, the instantaneous power injected at a bus $h$ can
be written as a complex quantity, as follows:
\begin{equation}
  \label{eq:s}
  \cp{s}_h = \cp{v}_h \, \cpc{\ii}_h \, ,
\end{equation}
where $^*$ denotes the complex conjugate operator. The time derivative
(rate of change) of complex power is:
\begin{equation}  
  \label{eq:sdot}
  \begin{aligned}
    \cpd{s}_h &= \cpd{v}_h \, \cpc{\ii}_h + \cp{v}_h \, \cpdc{\ii}_h \\
    &= \fq{v} \, \cp{v}_h  \, \cpc{\ii}_h + \cp{v}_h \, \fqc{\ii} \, \cpc{\ii}_h \\
    &= (\fq{v} + \fqc{\ii}) \, \cp{s}_h \, .
  \end{aligned}
\end{equation}
Using \eqref{eq:vdot2} and \eqref{eq:idot2}, \eqref{eq:sdot} can be
rewritten using the complex frequencies on the device local reference
frame:
\begin{equation}  
  \label{eq:sdot2}
  \begin{aligned}
    \cpd{s}_h
    &= \big [ \fq{v} + \jdd - \jdd + \fqc{\ii} \big ] \, \cp{s}_h \\
    &= \big [ \fq{v}' + (\fq{\ii}')^* \big ] \, \cp{s}_h \, .
  \end{aligned}
\end{equation}
Expression \eqref{eq:sdot} is general for any device connected to the
grid while expression \eqref{eq:sdot2} is general for any device that
has a synchronization mechanism that aligns the $dq$ reference frame
of the device with that of the grid.

\subsection{Ideal controllers}

The most common ideal models of devices connected to the grid are
considered in this section as they represent the operation of ideal
(i.e., perfect tracking and infinitely fast) generators and converter
controllers.  In the continuation, \eqref{eq:sdot} is coupled with the
characteristics of each device.  This allows to determine the impact
of these ideal devices on the \ac{cf} of the voltage and the current.

\subsubsection{Ideal Voltage Source (Slack Bus)}

An ideal slack bus has:
\begin{equation}  
  \label{eq:slack}
  \cpd{v}_h=0 \, , \quad \fq{v}=0 \, , 
\end{equation}
From \eqref{eq:slack}, \eqref{eq:sdot} becomes:
\begin{equation}  
  \label{eq:sdotslack}
  \begin{aligned}
    \cpd{s}_h
     &=\fqc{\ii} \, \cp{s}_h \, ,
  \end{aligned}
\end{equation}
which indicates that an ideal slack bus affects only the frequency of
the injected current, but not the voltage.

\subsubsection{Ideal Current Source}
An ideal current source that can impose both the magnitude and the
phase of the current:
\begin{equation}  
  \label{eq:slackiota}
  \cpd{\ii}_h=0 \, , \quad \fq{\ii}=0 \, ,
\end{equation}
leading to:
\begin{equation}  
  \label{eq:sdotcurrentslack}
  \begin{aligned}
    \cpd{s}_h
    &= \fq{v} \, \cp{s}_h \, .
  \end{aligned}
\end{equation}

\subsubsection{Current Source with Constant Power Factor}
More commonly, an ideal current source imposes the magnitude and a
constant power factor~\cite{freqcomplex}.  This results in:
\begin{equation}  
  \label{eq:Iphi1}
    \rho_{\ii}=0 \, , \quad \omega_v=\omega_{\ii} \, ,
\end{equation}
and thus:
\begin{equation}  
  \label{eq:Iphi22}
   \cpd{s}_h=\rho_v \cp{s}_h \, .
\end{equation}

\subsubsection{Constant Power}
For and ideal constant power load or source:
\begin{equation}  
  \label{eq:PQload}
    \cpd{s}_h=0 \, , \quad \fq{v} =- \fqc{\ii} \, , 
\end{equation}
which indicates that the complex frequencies of both voltage and
current change, but not independently.

\subsubsection{Constant Admittance}

For an ideal admittance, the magnitude of current is proportional to
the one of the voltage while their phases vary only by the constant
angle of the admittance.  This directly dictates
that~\cite{freqcomplex}:
\begin{equation}  
  \label{eq:Z1}
    \rho_v=\rho_{\ii} \, , \quad \omega_v=\omega_{\ii} \, ,
\end{equation}
and from \eqref{eq:sdot}:
\begin{equation}  
  \label{eq:Z2}
   \cpd{s}_h=2 \rho_v \cp{s}_h \, .
\end{equation}
This is consistent with the fact that an admittance cannot affect the
frequency at a bus.  Equations \eqref{eq:Iphi22} and \eqref{eq:Z2}
reveal that an ideal admittance and an ideal current source with
constant power factor have a similar dynamic response.

\subsubsection{PV Generator}

Lastly, an ideal PV generator has constant voltage magnitude and
constant active power:
\begin{equation}  
  \label{eq:PV1}
  \rho_v=0 \, , \quad \dot{p}=0 \, .
\end{equation}
Combining the two and substituting in \eqref{eq:sdot} leads to:
\begin{equation}  
  \label{eq:PV2}
    \frac{q}{p}=
    \frac{\rho_{\ii}}{(\omega_v-\omega_{\ii})} \, .
\end{equation}
where $p$ and $q$ are the active and reactive power components of
$\cp{s}_h$, respectively.  Equation \eqref{eq:PV2} leads to the
non-intuitive conclusion that the reactive power at a PV bus can be
changed not only by varying the voltage magnitude but also by varying
either the frequency of the voltage and/or of the current.
\begin{figure}
  \centering
  \includegraphics[width=0.51\columnwidth, height=4.2cm]{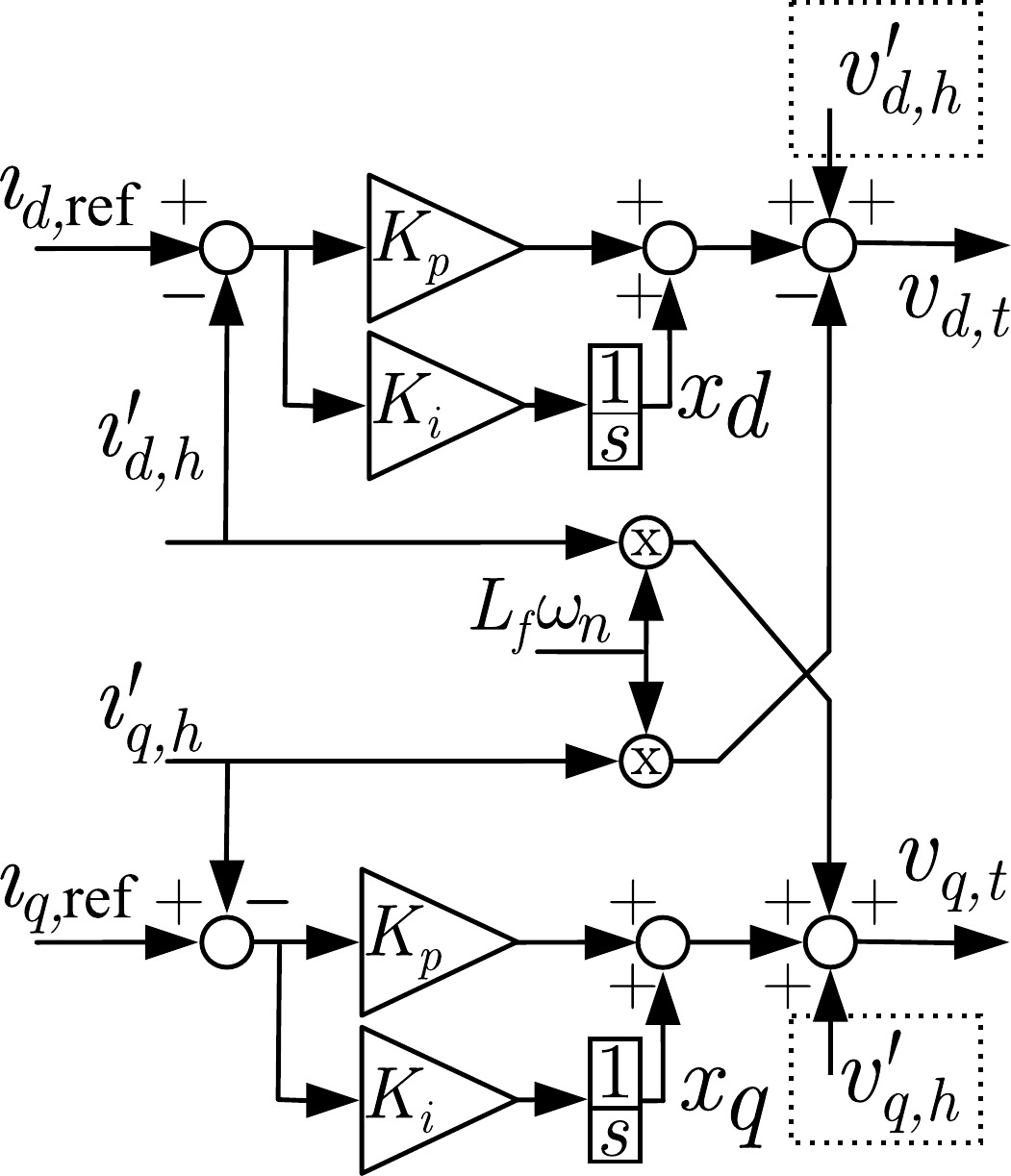}
  \caption{Control diagram of a PI-based current control in the
    synchronous reference frame. Optional \ac{vff} are specifically
    marked with dotted boxes.}
  \label{fig.currentcontrol}
  \vspace{-2mm}
\end{figure}

\subsection{Current Controller}

Due to practical considerations (e.g., over-current limiting and
resonance damping) an inner current controller is ubiquitous in VSC
applications \cite{wang2020grid}.  Figure \ref{fig.currentcontrol}
shows a control diagram of such a controller.  The dynamic equation of
a current flowing through the output inductor $L$ of a converter is:
\begin{equation}
  \label{eq:lvdrop}
  \cp{v}_t-\cp{v}_h = L_f \, \cpd{\ii}_h+R_f \, \cp{\ii}_h,
\end{equation}
where $L_f$ is the filter inductance, $R_f$ its parasitic resistance
and $\cp{v}_t=v_{d,t}+\jj v_{q,t}$ is the modulated voltage output of
the converter.  From Figure~\ref{fig.currentcontrol}, the control
equations can be derived as:
\begin{equation}
  \label{eq:vscout}
    \begin{aligned}
  \cp{v}_t &= \cp{v}'_h+\jj \omega_n L_f  \, \cp{\ii}'_h + K_p (\rf{\cp{\ii}}-\cp{\ii}'_h)+K_i \, \cp{x} \, , \\
  \cpd{x} &= \rf{\cp{\ii}} - \cp{\ii}_h'
    \end{aligned}
\end{equation}
where $K_p$, $K_i$ are the proportional and integral gains of the PI
controller, respectively, $\cp{x}=x_d+\jj x_q$ is the complex internal
state of the integral part of the controller and
$\rf{\cp{\ii}}= \ii_{d, \rm ref}+ \jj \ii_{q, \rm ref}$ is the current
reference signal.  By neglecting the synchronization
mismatches~($\cp{v}'_h \approx \cp{v}_h$,
$\cp{\ii}'_h \approx \cp{\ii}_h$), the electromagnetic dynamics of the
$L$-filter~($L_f \, \cpd{\ii}_h=L_f \, \fq{\ii} \, \cp{\ii}_h \approx
\omega_n L_f \, \cp{\ii}_h$) and its resistance~($R_f \approx 0$), and
by substituting equation \eqref{eq:vscout} to \eqref{eq:lvdrop}, one
derives~\cite{yazdani2010voltage}:
\begin{equation}  
  \label{eq:PI_L_dyn}
  \begin{aligned}
    0 &= K_p (\rf{\cp{\ii}} - \cp{\ii}_h') + K_i \, \cp{x} \, , \\
    \cpd{x} &= \rf{\cp{\ii}} - \cp{\ii}_h' \, .
    \end{aligned}
\end{equation}
Differentiating the algebraic equation in \eqref{eq:PI_L_dyn}, one
has:
\begin{equation}  
  \label{eq:PI_constraint_derivative}
  \begin{aligned}
    \fq{\ii}' \, \cp{\ii}_h' &= \rf{\cpd{\ii}} + \kpi \, \cpd{x} \\
    &= \fq{\rf{\ii}} \rf{\cp{\ii}} + \kpi \, (\rf{\cp{\ii}} - \cp{\ii}_h') \\
    &= (\fq{\rf{\ii}} + \kpi{}) \, \rf{\cp{\ii}} - \kpi{} \, \cp{\ii}_h' \, .
  \end{aligned}
\end{equation}
where $\kpi = K_i/K_p$.  Equation \eqref{eq:PI_constraint_derivative}
can be also written as:
\begin{equation}  
  \label{eq:PI_constraint_derivative_2}
  ( \fq{\ii}' + \kpi{}) \, \cp{\ii}_h' =
  ( \fq{\rf{\ii}} + \kpi{}) \, \rf{\cp{\ii}} \, ,
\end{equation}
which shows that the coefficient $\kpi{}$, and hence the PI controller
effect, can be interpreted as a constant real translation that affects
only the magnitude.  Then, substituting
\eqref{eq:PI_constraint_derivative} into \eqref{eq:sdot2} leads to:
\begin{equation}  
  \label{eq:sdot3}
  \cpd{s}_h = ( \fqc{\rf{\ii}} + \kpi{}) \, \cp{v}'_h \,
  \rf{\cpc{\ii}} + (\fq{v}' - \kpi{}) \, \cp{s}_h \, .
\end{equation}
Equation \eqref{eq:sdot3} separates the effect of the inner current
control from the synchronization mechanism and the outer control
loops.  Moreover, if the dynamic of the PI can be assumed fast,
$\rf{\cp{\ii}} \approx \cp{\ii}_h'$ and hence
$\fq{\rf{\ii}} \approx \fq{\ii}'$.  Then \eqref{eq:sdot3} simplifies
to \eqref{eq:sdot2}.  Finally we note that the coefficients of
$\cp{\ii}'$ and $\cp{v}'$ that appears in
\eqref{eq:PI_constraint_derivative_2} and \eqref{eq:sdot3},
respectively, can be written as:
\begin{equation}
  \begin{aligned}
    (\fq{\ii}')^* + \kpi{} &= \fqc{\ii} + ( \kpi{} + \jdd{} ) \, , \\
    \fq{v}' - \kpi{} &= \fq{v} - (\kpi{} + \jdd{}) \, .
  \end{aligned}
\end{equation}
The complex quantity $\kpi{} + \jdd{}$ has the dimension of a complex
frequency and embeds the effects of the synchronization mechanism and
of the current control.  The equation is general and can be applied to
both synchronous machines and controllers with an internal current
loop.  For a synchronous machine $\kpi{} = 0$, whereas for an ideal
synchronization (e.g., ideal \ac{pll}), $\jdd \approx 0$ because
$\delta \approx 0$.  The utilization of the \ac{cf} allows thus to
easily identify and separate the dynamic effects and non-ideality of
the controllers of conventional and converter-interfaced devices
connected to the grid.

\subsection{Voltage Feed-Forward (VFF)}
Equation \eqref{eq:PI_L_dyn} assumes that a \ac{vff} is included in
the current controller \cite{yazdani2010voltage}.  If the \ac{vff} is
not included, \eqref{eq:PI_L_dyn} becomes:
\begin{equation}  
  \label{eq:PI_constraint2}
  \cp{\ii}_h' = \rf{\cp{\ii}} + \frac{1}{K_p}(K_i \cp{x} - \cp{v}'_h) \, ,
\end{equation}  
which leads to:
\begin{equation}
  \label{eq:idotvff2}
  ( \fq{\ii}' + \kpi{}) \, \cp{\ii}_h' =
  ( \fq{\rf{\ii}} + \kpi{}) \, \rf{\cp{\ii}}
  - \frac{1}{K_p} \fq{v}' \, \cp{v}'_h \, ,
\end{equation}
and, finally:
\begin{equation}
  \label{eq:sdotvff2}
   \cpd{s}_h = ( \fqc{\rf{\ii}} + \kpi{}) \, \cp{v}'_h \,
  \rf{\cpc{\ii}} + (\fq{v}' - \kpi{}) \, \cp{s}_h \, - \frac{1}{K_p} \, (\fq{v}')^* v_h^2 \, ,
\end{equation}
where $v_h=v_h'$ is the magnitude of $\cp{v}'_h$.  This result
indicates that the dynamic effect of the \ac{vff} is inversely
proportional to $K_p$ and that it modulates the \ac{cf} of voltage
$\cp{v}_h$.

\begin{figure}
  \centering
  \includegraphics[width=0.475\columnwidth, height=1.45cm]{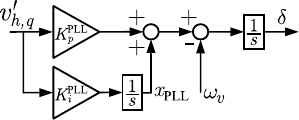}
  \caption{Control diagram of synchronous reference frame \ac{pll}.}
  \label{fig.pll}
  \vspace{-2mm}
\end{figure}

\section{\ac{gfl} Converters}
\label{sec.gfl}

\ac{gfl} converters typically measure or estimate the frequency and
phase of the voltage at the converter connection point and then use
these measurements to synchronize with the grid~\cite{wang2020grid}.
Typically, this measurement is achieved through a \ac{pll} device.  A
common \ac{pll} synchronization strategy is to set the voltage
$q$~component equal to zero by using a PI controller.  In that case,
the output of the \ac{pll} provides the angle that is used for the
transformation between converter-side and grid-side variables.  A
control diagram of a common \ac{pll} is shown in Fig.~\ref{fig.pll}.
The synchronization strategy is given by \cite{ortega2018comparison}:
\begin{equation}  
  \label{eq:PLL}
  \begin{aligned}
    &\Dot{x}_{\pll}
    =
    \Im \{ \cp{v}_h' \}
    \, ,
    \\
    &\Dot{\delta}
    =
    K_p^{\pll} \Im \{ \cp{v}_h' \}
    +
    K_i^{\pll} x_{\pll}
    -
    \omega_v
    \, ,
  \end{aligned}
\end{equation}
where $K_p^{\pll}$, $K_i^{\pll}$ are the proportional and integral
gains of the PI controller included in the \ac{pll}, $\Im$ is the
imaginary part operator and $\omega_v$ is the frequency at the
connection bus.  Many other \ac{pll} implementations are reported in
the literature~\cite{ortega2018comparison,ali2018three}.
Equation~\eqref{eq:PLL} can be substituted back into
equations~\eqref{eq:vdot2} and~\eqref{eq:idot2} so that the effect of
the \ac{pll} parameters to the \ac{cf} can be directly evaluated.

In the following subsections, expressions for the terms
$\rf{\cp{\ii}}$ and $\fq{\rf{\ii}}$~(or
$\fq{\rf{\ii}} \, \rf{\cp{\ii}}$) are provided for relevant \ac{gfl}
control configurations that are typically found in practical
applications.  Then, these expressions are coupled with
\eqref{eq:sdot3} and \eqref{eq:PLL} to derive the expression of
$\cpd{s}_h$.

\subsection{Current controller with constant current reference}
In this simplified case, the current controller tracks a constant
current reference.  Thus, the expressions for $\rf{\cp{\ii}}$ and
$\fq{\rf{\ii}}$ are immediately deduced as:
\begin{equation}  
  \label{eq:current_reference}
  \begin{aligned}
    \rf{\cp{\ii}}
    &=
    \rm const.
    \, ,
    \\
    \Rightarrow \quad \fq{\rf{\ii}}
    &=
    0
    \, ,
  \end{aligned}
\end{equation}
and substituting into \eqref{eq:sdot3}:
\begin{equation}  
  \label{eq:sdot_ccc}
  \cpd{s}_h =  \kpi{} \, \cp{v}'_h \,
  \rf{\cpc{\ii}} + (\fq{v}' - \kpi{}) \, \cp{s}_h \, .
\end{equation}
For an ideal current source, $\rf{\cp{\ii}}=\cp{\ii}$ and substituting
$\cp{v}'_h \rf{\cpc{\ii}}=\cp{s}_h$, \eqref{eq:sdot_ccc} simplifies to
\eqref{eq:sdotcurrentslack}.  With the use of \eqref{eq:PI_L_dyn},
equation~\eqref{eq:sdot_ccc} can be re-written in terms of the
controller internal states as:
\begin{equation}  
  \label{eq:sdot_ccc2}
  \cpd{s}_h =  \cp{s}_h \fq{v}- \kpi{}^2 \cpc{x} \cp{v}_h'
  = \cp{s}_h \fq{v}+ \Delta \cpd{s}_h
  \, .
\end{equation}
In the following, \eqref{eq:sdot_ccc2} is utilized to highlight the
effect of the non-ideal current controller in comparison with the
ideal current source of \eqref{eq:sdotcurrentslack}.

\begin{figure}
  \centering
  \includegraphics[width=0.75\columnwidth, height=2.3cm]{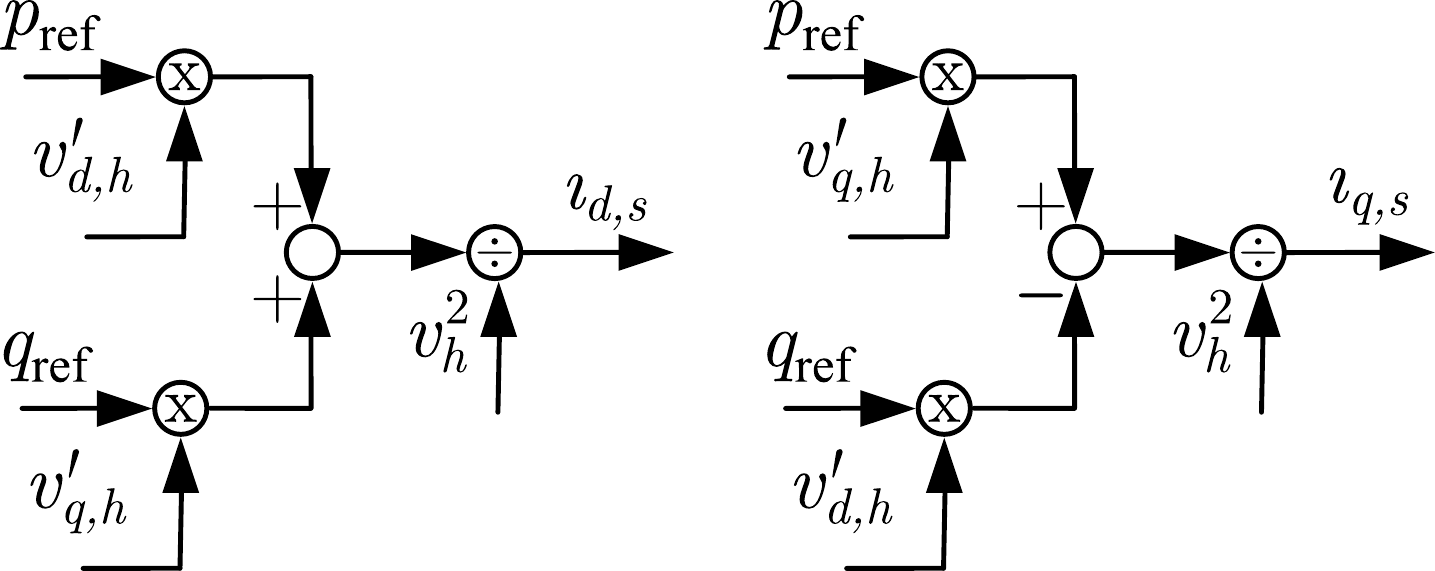}
  \caption{Block diagram for the calculation of current reference from
    active and reactive power commands.}
  \label{fig.powerrefcalc}
  \vspace{-2mm}
\end{figure}

\subsection{Current Controller with Constant Power Reference}

In practical applications, a constant power reference ($\rf{\cp{s}}$)
is often preferred over a current reference~\cite{teodorescu2011grid}.
Then:
\begin{equation}  
  \label{eq:power_reference}
  \begin{aligned}
    \rf{\cp{\ii}}
    &=
    \cp{\ii}_s
    =
    \left(\frac{\rf{\cp{s}}
      }{\cp{v}_h'}\right)^*
    \, ,
    \\
    \Rightarrow \quad
    \fq{\rf{\ii}} \, \rf{\cp{\ii}}
    &=
    \left(-\frac{\rf{\cp{s}}}{\cp{v}_h'}\fq{v}'\right)^*
    \\
    &=
    -\rf{\cp{\ii}} \, (\fq{v}')^* \, ,
    \\
    \Rightarrow \quad 
    \fq{\rf{\ii}}
    &=
    -(\fq{v}')^* \, ,
  \end{aligned}
\end{equation}
leading to:
\begin{equation}  
  \label{eq:sdot_cpcc}
  \cpd{s}_h =
  (\fq{v}'-\kpi{}) \, (\cp{s}_h-\rf{\cp{s}}) \, .
\end{equation}
For an ideal PQ source $\cp{s}_h=\rf{\cp{s}}$, which leads to simplify
\eqref{eq:sdot_cpcc} to \eqref{eq:PQload}.  A block diagram for the
calculation of the current reference from the active and reactive
power commands is shown in Fig.~\ref{fig.powerrefcalc}.

\begin{figure}
  \centering
  \includegraphics[width=0.555\columnwidth, height=4.1cm]{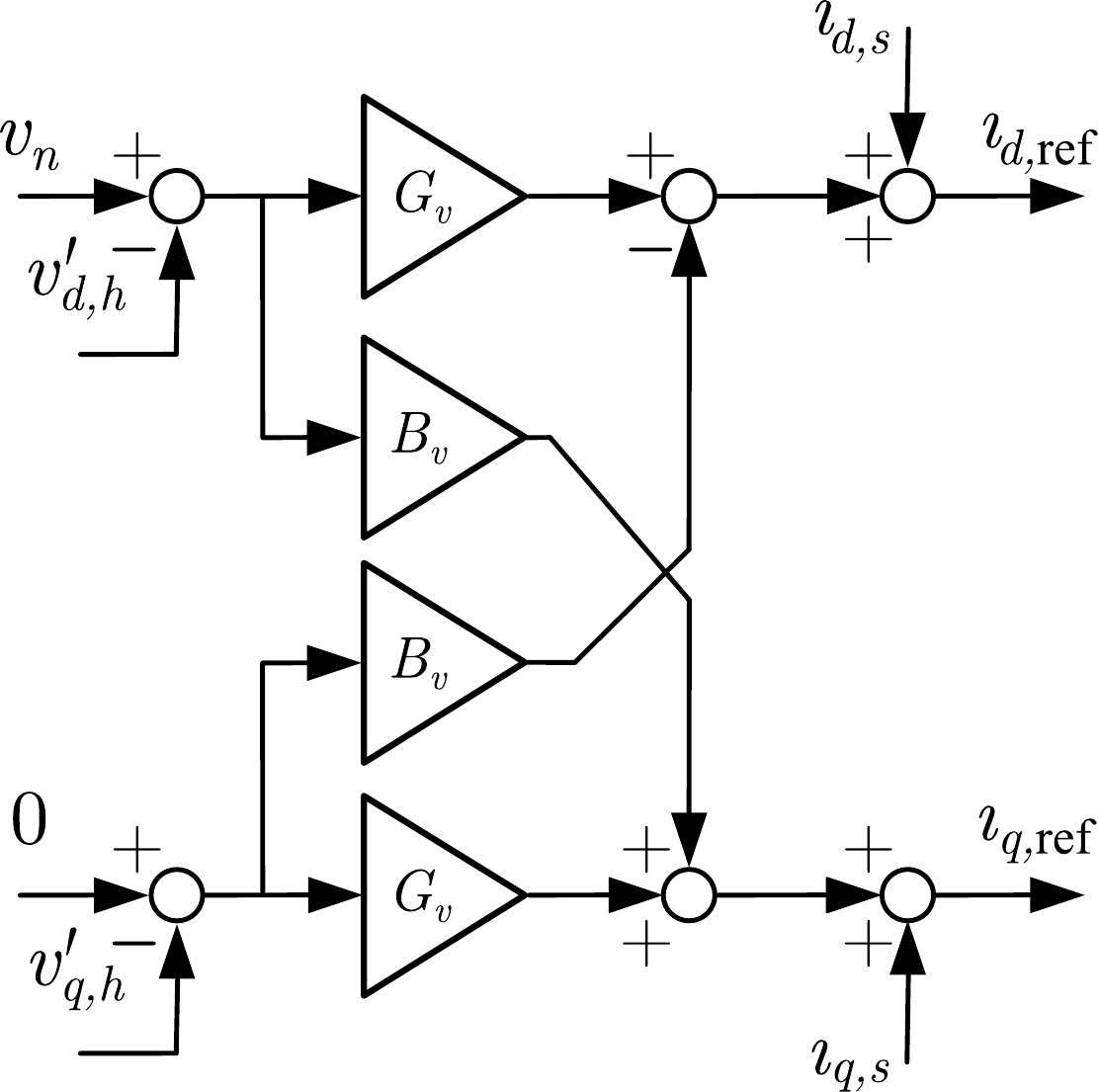}
  \caption{Control diagram of a virtual admittance loop.}
  \label{fig.valscheme}
  \vspace{-2mm}
\end{figure}

\subsection{Current Controller with Virtual Admittance Loop}

Reference \cite{moutevelis2021virtual} proposes a virtual admittance
control loop able to provide voltage support in resistive or weak
grids.
A control diagram of this loop can be seen in Fig.~\ref{fig.valscheme}.
This loop emulates the operation of an admittance $\cp{Y}_v=G_v+\jj B_v$ in order to follow a voltage reference $\rf{\cp{v}}=v_n+\jj 0$, where $v_n$ is the nominal voltage~($\rf{\cp{v}}=\rf{\cp{v}}^*$).
In the literature, converters that provide ancillary services to the grid are also referred to as grid-supporting converters~\cite{rocabert2012control}.
However, since these controllers utilizes identical \ac{pll} and current controllers, they can be included in the \ac{gfl} category for ease of reference.
The equations are:
\begin{equation}  
\label{eq:virtual_admittance}
\begin{aligned}
\rf{\cp{\ii}}
&=
\cp{\ii}_s
+
\overline{Y}_v (\rf{\cp{v}}-\cp{v}_h')
\\
&=
(\frac{\rf{\cp{s}}
}{\cp{v}_h'})^*
+
\overline{Y}_v \left(\rf{\cp{v}}-\cp{v}_h'\right)
\, ,
\\
\Rightarrow \quad \fq{\rf{\ii}} \rf{\cp{\ii}}
&=
\left(-\frac{\rf{\cp{s}}}{\cp{v}_h'}\fq{v}'\right)^*
-
\overline{Y}_v \cp{v}_h' \fq{v}'
\, ,
\end{aligned}
\end{equation}
where $\cp{\ii}_s$ denotes the current calculated by the power reference as in the previous case. By substituting:
\begin{equation}  
  \begin{aligned}
  \label{eq:sdot_val}
  \cpd{s}_h 
  &=
  (\fq{v}'-\kpi{}) \, (\cp{s}_h-\rf{\cp{s}})
\\
  &+
  \cpc{Y}_v (-v_h^2 ((\fq{v}')^*+\kpi{})+\kpi{} \cp{v}_h' \rf{\cp{v}})
  \, .
  \end{aligned}
\end{equation}
Assuming ideal current control~($\kpi{}=0$), \eqref{eq:sdot_val} simplifies to:
\begin{equation}  
  \begin{aligned}
  \label{eq:sdot_val_simple}
  \cpd{s}_h 
  &=
  -v_h^2 \cpc{Y}_v 2 \rho_v + \cpc{Y}_v \fq{v}' \cp{v}_h' \rf{\cp{v}}
  \, .
  \end{aligned}
\end{equation}
The first term corresponds to the power consumed by the virtual admittance and is consistent with \eqref{eq:Z2}, whereas the second term corresponds to the voltage reference.
%
%
\begin{figure}
\centering
\includegraphics[width=0.5\columnwidth, height=3cm]{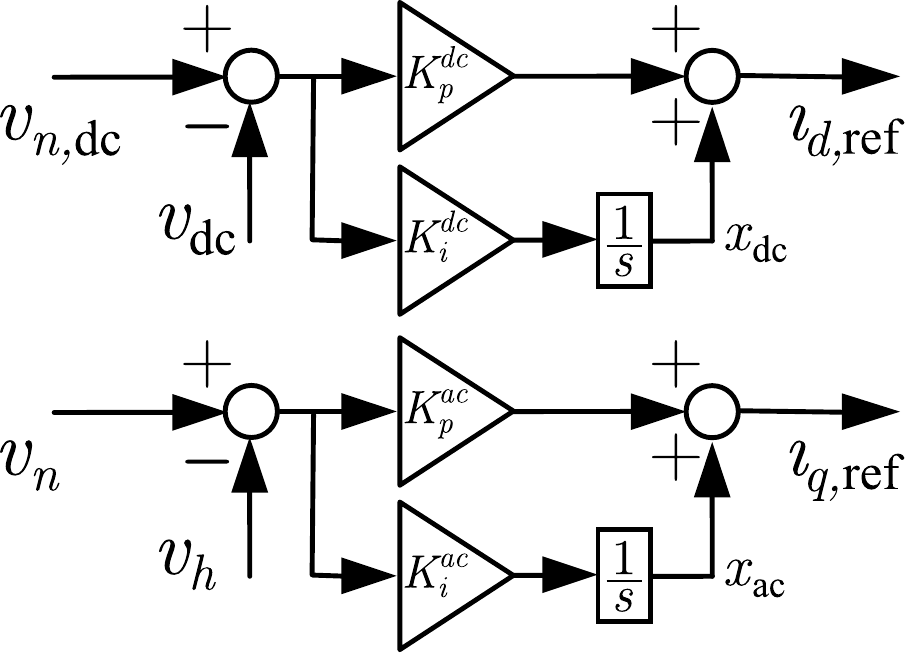}
\caption{Control diagram of the outer loops for \ac{gfl} applications.}
\label{fig.gflouterscheme}
\end{figure}
\subsection{Current Controller with Outer Control Loops}
\ac{gfl} converters can be used for  applications different than power delivery from distributed sources~\cite{teodorescu2011grid}.
Most notable is the case of \ac{gfl} rectifiers that are tasked with maintaining dc-link voltage $v_{\rm dc}$ equal to its nominal value $v_{n, \rm dc}$ of an electronically-interfaced load~\cite{yang2022bifurcations}.
This is achieved through an outer-loop that calculates the $d$-axis current reference~\cite{yazdani2010voltage,huang2017bifurcation}.
Therefore, the $q$-axis current is free to provide voltage support.
This is usually achieved through an outer droop controller but a PI regulator can also be used~\cite{vasquez2009voltage,yang2020nonlinear}.
In this study, a PI is used for generality. 
A control diagram of these outer loops can be seen in Fig.~\ref{fig.gflouterscheme}.

%
The current references are calculated as:
\begin{equation}  
\label{eq:outer_loop_grid_following}
\begin{aligned}
\iota_{d,\rm ref}
&=
K_p^{\rm dc}(v_{n, \rm dc}-v_{\rm dc})
+
K_i^{\rm dc} x_{\rm dc}
\, ,
\\
\iota_{q,\rm ref}
&=
K_p^{\rm ac}(v_n-v_h)
+
K_i^{\rm ac} x_{\rm ac}
\, ,
\end{aligned}
\end{equation}
where $K_p^{\rm dc}$, $ K_i^{\rm dc}$ are the proportional and
integral gains, respectively, of the outer $\rm dc$ voltage PI
controller, $K_p^{\rm ac}$, $ K_i^{\rm ac}$ are the proportional and
integral gains, respectively, of the outer $\rm ac$ voltage PI
controller, $v_n$ is the nominal voltage rate of the connection bus,
and the internal PI states are:
\begin{equation}  
\label{eq:outer_loop_PI_states}
\begin{aligned}
\dot{x}_{\rm dc}
&=
v_{n, \rm dc}-v_{\rm dc}
\, ,
\\
\dot{x}_{\rm ac}
&=
v_n-v_h
\, .
\end{aligned}
\end{equation}
By using the definition of the real part of the \ac{cf} \eqref{eq:udot}, equation \eqref{eq:outer_loop_grid_following} and its derivative can be written using complex notation as:
\begin{equation}  
\label{eq:outer_loop_complex}
\begin{aligned}
\rf{\cp{\ii}}
&=
K_p^o (\rf{\cp{v}}^o-\cp{v}_o)+K_i^o\cp{x}_o
\, ,
\\
\Rightarrow \quad 
\fq{\rf{\ii}} \, \rf{\cp{\ii}}
&=-\cp{v}_o(K_p^o\cp{\rho}+K_i^o)+K_i^o \rf{\cp{v}}^o \, ,
\end{aligned}
\end{equation}
where $K_p^o=K_p^{\rm dc}=K_p^{\rm ac}$, $K_i^o=K_i^{\rm dc}=K_i^{\rm ac}$, $\cp{x}_o=x_{\rm dc}+\jj x_{\rm ac}$, $\rf{\cp{v}}^o=v_{n, \rm dc}+\jj v_n$, $\cp{v}_o=v_{\rm dc}+v_h$ and $\cp{\rho}=\rho_{\rm dc}+\jj \rho_v$.
The assumption of equal controller gains for the two $dq$-axes is consistent if a proper pu base for the dc voltage is selected so that $v_{n, \rm dc}=v_{n}$.
This is straightforward as the dc voltage is independent from the rest of the system and typically ac and dc voltages are selected in the same order of magnitude~\cite{yazdani2010voltage}.
Equation \eqref{eq:outer_loop_complex} indicates that the outer control loops only affect quantity $\cp{\rho}$ (i.e., the magnitude of ac and dc voltage) but not frequency $\omega_v$.
%
\section{\ac{gfm} Converters}
\label{sec.gfm}
%
%
In this section, the analytical expressions for $\rf{\cp{\ii}}$, $\fq{\rf{\ii}} \rf{\cp{\ii}}$ and $\cpd{s}$ are derived, considering the inner voltage controllers for \ac{gfm} control configurations.
Then, different power-based synchronization and voltage regulation strategies that are commonly found in the literature are presented. 
\begin{figure}
\centering
\includegraphics[width=0.5\columnwidth, height=3cm]{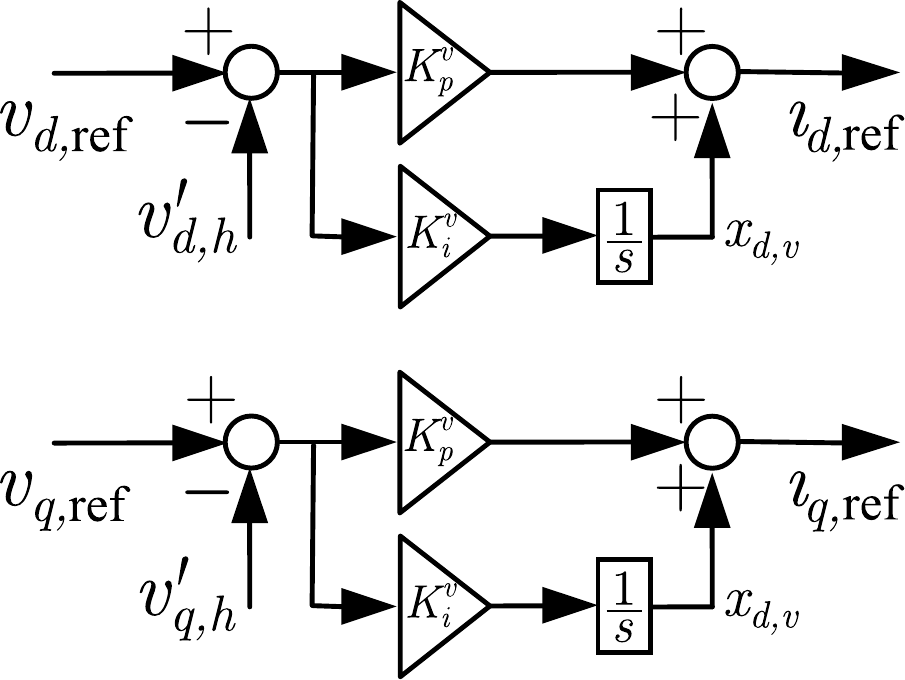}
\caption{Control diagram of a PI-based voltage controller used in \ac{gfm} applications.}
\label{fig.gfmvoltage}
\vspace{-2mm}
\end{figure}
\subsection{Voltage Controller}
A set of equations that describes a PI-based, voltage controller is the following:
\begin{equation}  
\label{eq:simple_voltage_controller}
\begin{aligned}
\cpd{x}_v
&=
\rf{\cp{v}}
-
\cp{v}_h'
,
\\
\rf{\cp{\ii}}
&=
K_{p}^v (\rf{\cp{v}}
-
\cp{v}_h'
)
+
K_{i}^v \overline{x}_v
,
\\
\fq{\rf{\ii}} \rf{\cp{\ii}}
&=
(K_p^v \fq{\rf{v}}+K_i^v) \rf{\cp{v}}
-
(K_p^v \fq{v}'+K_i^v) \cp{v}_h'
.
\end{aligned}
\end{equation}
where $\rf{\cp{v}}$ is the voltage reference from an outer loop, e.g., a voltage droop controller.
The control diagram of the outer voltage controller for \ac{gfm} applications is shown in Fig.~\ref{fig.gfmvoltage}.
To decouple the effects of the voltage and the current controllers, it is assumed that $\fqc{\rf{\ii}} + \kpi{} \approx \fqc{\rf{\ii}}$. Thus, by substituting the third equation of \eqref{eq:simple_voltage_controller} into \eqref{eq:sdot3}, one has:
\begin{equation}  
  \label{eq:sdot_simplev}
  \begin{aligned}
  \cpd{s}_h =& \,
  (K_p^v \fqc{\rf{v}}+K_i^v) \cp{v}_h' \rf{\cpc{v}} 
  \\ &- (K_p^v (\fq{v}')^*+K_i^v) v_h^2 
  +
  (\fq{v}' - \kpi{}) \, \cp{s}_h 
  \, .
  \end{aligned}
\end{equation}
The above equation indicates that gain $K_p^v$ modulates the frequency of the voltage in a similar way as gain $K_p^o$ modulates the quantity $\cp{\rho}$ in \eqref{eq:outer_loop_complex}.
\begin{figure}
\centering
\includegraphics[width=0.675\columnwidth]{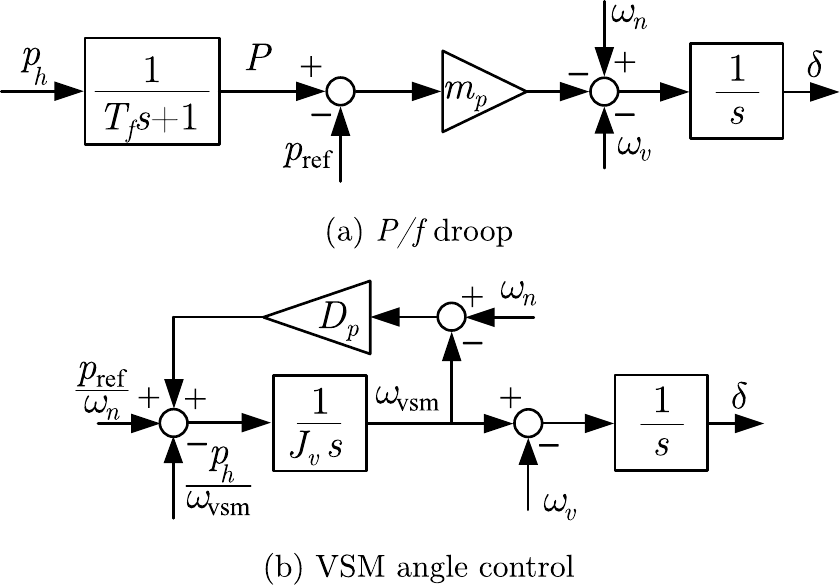}
\caption{Control diagram of the active power synchronization strategies. (a) $P/f$ droop and (b) \ac{vsm} angle control.}
\label{fig.activesyncscheme}
\vspace{-2mm}
\end{figure}
\vspace{-2mm}
\subsection{Power-Based Synchronization Strategies}
This section discusses the expressions for $\Dot{\delta}$ based on the
widely known active power-frequency droop~($P/f$ droop) and \ac{vsm}
concepts.
Control diagrams for these two synchronization strategies can be seen in Fig.~\ref{fig.activesyncscheme}.
Various other implementations for the outer frequency/phase controllers have been proposed and the optimal controller selection for different scenarios is still an open research topic~\cite{9408354}.

The equations for the $P/f$ droop are the following~\cite{4118327,moran2021influence}:
\begin{equation}  
\label{eq:pf_droop}
\begin{aligned}
\Dot{\delta}
&=
\omega_n
-
\omega_v
-
m_p (P-\rf{p})
\, ,
\\
T_f \Dot{P}
&=
p_h-P
\, ,
\end{aligned}
\end{equation}
where $m_p$ is the droop gain and $T_f$ is the time constant of the (optional) low-pass filter.
%
Lowercase $p_h$ is the active power injection and uppercase $P$ is the filtered active power.
For the \ac{vsm}, a common implementation in the literature is the
following~\cite{zhong2010synchronverters, roldan2019design,
  gonzalez2021design}:
\begin{equation}  
\label{eq:vsm}
\begin{aligned}
&\Dot{\delta}
=
\omega_{\vsm}-\omega_v
\, ,
\\
&\Dot{\omega}_{\vsm}
=
\frac{1}{J_v}
(
\frac{\rf{p}}{\omega_n}
-
\frac{p_h}{\omega_{\vsm}}
+
D_p (\omega_n-\omega_{\vsm})
)
\, ,
\end{aligned}
\end{equation}
where $J_v$ is the moment of virtual inertia coefficient while $D_p$
is the virtual damping coefficient.
It should be noted that moment of inertia $J$ is a distinct quantity
than the mechanical starting time $M$ and only coincide when pu values
are being used.
\begin{figure}
\centering
\includegraphics[width=0.65\columnwidth]{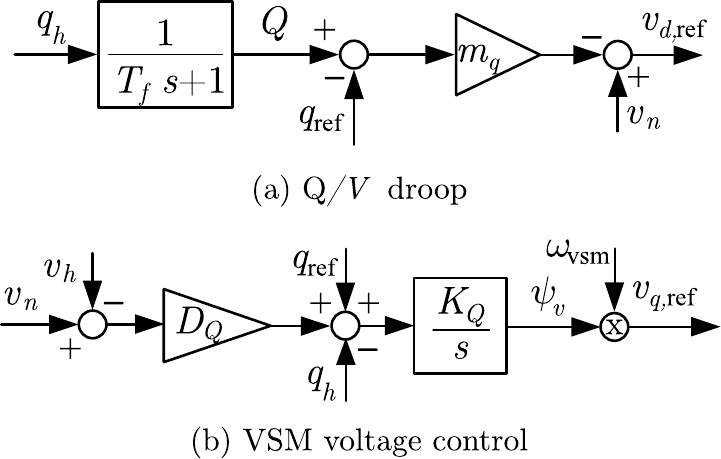}
\caption{Control diagram of the outer voltage loops for \ac{gfm} applications. (a) $Q/V$ droop and (b) \ac{vsm} voltage control.}
\label{fig.gfmvoltageouter}
\vspace{-2mm}
\end{figure}
\vspace{-2mm}
\subsection{Outer Voltage Loop}
This section discusses the outer voltage loop that defines the quantity $\rf{\cp{v}}=v_{d, \rm ref}+\jj v_{q, \rm ref}$.  By obtaining the analytical expressions for $\rf{\cp{v}}$ and $\cpd{v}_{\rm ref}$, they can be substituted into \eqref{eq:sdot_simplev}.
Typically, reactive power/voltage ($Q/V$) droop controllers, PI/integral-based controllers or a combination of the two are preferred~\cite{9408354}.
The control diagrams for the voltage regulation loops for the $Q/V$ droop strategy and the \ac{vsm} voltage control strategy are shown in Fig.~\ref{fig.gfmvoltageouter}.

For the droop controller, the voltage is aligned with the $d$-axis of the voltage. That results in:
\begin{equation}  
\label{eq:qv_droop}
\begin{aligned}
\rf{\cp{v}}
=
v_{d, \rm ref}
&=
v_n
-
m_q (Q-q_{ref})
\, ,
\\
T_f \Dot{Q}
&=
q_h-Q
\, ,
\end{aligned}
\end{equation}
where lowercase $q$ is the reactive power injection and uppercase $Q$ is the filtered reactive power.
Differentiating one gets:
\begin{equation}  
\label{eq:qv_droop_derivative}
\begin{aligned}
\fq{\rf{v}} \rf{\cp{v}}
=
\Dot{v}_{{d, \rm ref}}
&=
\frac{m_q}{T_f} (Q-q_h)
.
\end{aligned}
\end{equation}
By observing that $\rf{\cp{v}}=v_{d, \rm ref}$ leads directly to $\fq{\rf{v}} \rf{\cp{v}}=(\fq{\rf{v}} \rf{\cp{v}})^*$, one can simplify \eqref{eq:qv_droop_derivative} to:
\begin{equation}  
\label{eq:qv_droop_derivative2}
\begin{aligned}
\fq{\rf{v}} \rf{\cp{v}}
=
\rho_{\rf{v}} v_{d, \rm ref}
&=
\frac{m_q}{T_f} (Q-q_h)
.
\end{aligned}
\end{equation}
An example of a combination of integral and droop controller for the
voltage regulation can be found in~\cite{gonzalez2021design}.
This strategy is paired with the \ac{vsm} synchronization control of
equation \eqref{eq:vsm}.
In it, the equivalence with physical quantities is preserved by
defining the virtual flux as:
\begin{equation}  
\label{eq:virtual_flux}
\begin{aligned}
\Dot{\psi}_v
=
K_Q
(\rf{q}
-
q_h
+
D_Q (v_n
-
v_h
)
)
,
\end{aligned}
\end{equation}
where $K_Q$ is the integral gain and $D_Q$ is the droop gain.
The output of the outer voltage controller is then aligned with the
$q$-axis of the voltage as in:
\begin{equation}  
  \label{eq:v_q_VSM}
  \begin{aligned}
    &\rf{\cp{v}}
    =
    \jj v_{q, \rm ref}
    =
    \jj \psi_v \omega_{\vsm} \, ,
    \\
    \Rightarrow \quad
    & \fq{\rf{v}} \rf{\cp{v}}
    =
    \jj (\Dot{\psi}_v \omega_{\vsm}
    +
    \psi_v \Dot{\omega}_{\vsm}
    )
    \, ,
  \end{aligned}
\end{equation}
where $\Dot{\psi}_v$ is given by \eqref{eq:virtual_flux} and
$\Dot{\omega}_{\vsm}$ is given by \eqref{eq:vsm}.
Similar to the previous case, one observes that again
$\fq{\rf{v}}=\rho_{\rf{v}}$.
It can be concluded then that when either the real or the imaginary
part of the outer loop reference is set to zero, the outer loop does
not modify the frequency of the reference signal, only its magnitude.
This is consistent with the original design of the outer voltage
loops.

For ease of reference, all the internal frequencies derived in
Sections \ref{sec.gfl} and \ref{sec.gfm} are summarized in
Table~\ref{tab.internalfreq}.
Once these internal frequencies are defined, one can compare the
dynamic performance of converter control based on the bus frequency
and on the \textit{internal} ones, and also understand how the
parameters of the these controllers affect the internal frequency and,
then, again, the control itself.  This study is particularly relevant
in the case of electronic converters as their controllers can be fast
and thus even relatively small differences in the value of the
measured frequency can lead to visible differences in their dynamic
performance.
\begin{table}[t]
  \centering
  \caption{Summary of the internal frequency expressions}
  \vspace{-0.1cm}
  \footnotesize
  \renewcommand{\arraystretch}{1.15}
  \begin{tabular}{l|l|l}
    \hline
    Control Configuration & Internal Frequency & Equations \\
    \hline
    Current control + \ac{pll} & $\fq{v} - \jdd{} - \kpi{}$ & \eqref{eq:vdot2}, \eqref{eq:PLL}, \eqref{eq:sdot_ccc}, \eqref{eq:sdot_cpcc}   \\
    Effect of \ac{vff}& $- \frac{1}{K_p}(\fq{v}^* + \jdd{})$ & \eqref{eq:vdot2}, \eqref{eq:sdotvff2}, \eqref{eq:PLL}\\
    Virtual admittance & $ -v_h^2 \cpc{Y}_v ((\fq{v}')^*+\kpi{})$ & \eqref{eq:vdot2}, \eqref{eq:PLL}, \eqref{eq:sdot_val}\\
    \ac{gfl} outer loops& $K_p^o\cp{\rho}+K_i^o$ & \eqref{eq:outer_loop_complex}\\
    \ac{gfm} voltage control & $-
                               (K_p^v (\fq{v}')^*+K_i^v)$ & \eqref{eq:sdot_simplev}\\
    $P/f$ droop control & $\fq{v} - \jdd{}$ &\eqref{eq:vdot2}, \eqref{eq:pf_droop}\\
    \ac{vsm} angle control & $\fq{v} - \jdd{}$ &\eqref{eq:vdot2}, \eqref{eq:vsm}\\
    \hline
  \end{tabular}
  \vspace{+0.2cm}
  \label{tab.internalfreq}
\end{table}
\begin{figure}[htb]
  \centering
  \includegraphics[width=0.65\columnwidth]{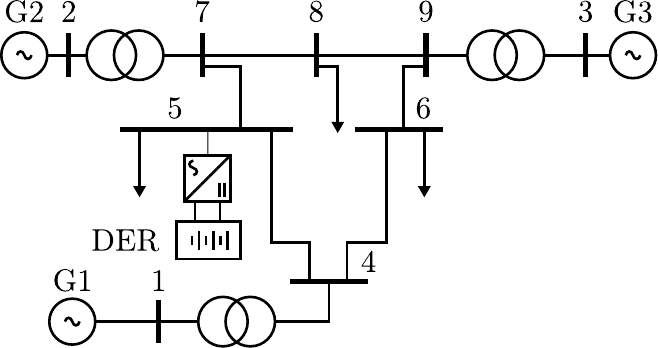}
  \caption{Single-line diagram of the modified WSCC 9-bus system.}
  \label{fig.wscc}
  \vspace{-2mm}
\end{figure}

\section{Simulation Results}
\label{sec.simresults}
To illustrate the theoretical developments, different case studies
considering the WSCC 9-bus system are presented in this section.
The purpose of the case study is to show how the \ac{cf} can be used
as a \textit{metric} for the dynamic performance of the
synchronization mechanisms and controllers of grid-connected
converters.
The synchronous generators are modeled using conventional 4th order
models.
A converter-interfaced energy storage system is connected to bus 5 of
the grid through an $LC$-filter, modeled as
in~\cite{ortega2015generalized}.
Although its dynamic effects has been neglected in the analytical
developments above, the ac filter is included in the models utilized
in the simulations.
A schematic of the network that was used for all simulations is shown
in Figure~\ref{fig.wscc}.
The \ac{cf} components $\rho_v$ and $\omega_v$ are calculated using
the procedure described in~\cite{freqcomplex}.
All simulations were carried out with the Dome software
tool~\cite{milano2013python}.
For the examples regarding the effect of the \ac{pll} and the current
control, a standard SRF-\ac{pll} described by \eqref{eq:PLL} is used.
The current references are kept constant and tracked by a PI
controller as in \eqref{eq:PI_L_dyn}.
The dc voltage is kept constant by an ideal voltage source.  For the
example on the outer controllers, the ideal voltage source is swapped
for a capacitor of $C_{\rm dc} = 0.05$~F and the outer loops of
\eqref{eq:outer_loop_grid_following} and
\eqref{eq:outer_loop_PI_states} are used.
Lastly, for the \ac{gfm} configurations, the outer voltage loop of
\eqref{eq:simple_voltage_controller} are used to regulate the voltage.
Equations \eqref{eq:pf_droop} and \eqref{eq:qv_droop} are used to
define the droop-based, \ac{gfm} control while \eqref{eq:vsm},
\eqref{eq:virtual_flux} and \eqref{eq:v_q_VSM} are used for the
\ac{vsm} control.  In the latter case, an ideal voltage source is
again used to regulate the dc voltage.
The contingency that is used for all examples is a disconnection of
the load at bus 5 at $t=1$~s.
If not explicitly stated otherwise, all frequency-like quantities in
this section refer to frequency deviations from the nominal
values~($\rho_v=0$ and $\omega_v=1$ pu).

\subsection{Effect of the \ac{pll}}
\begin{figure}
\centering
\includegraphics[width=0.49\columnwidth]{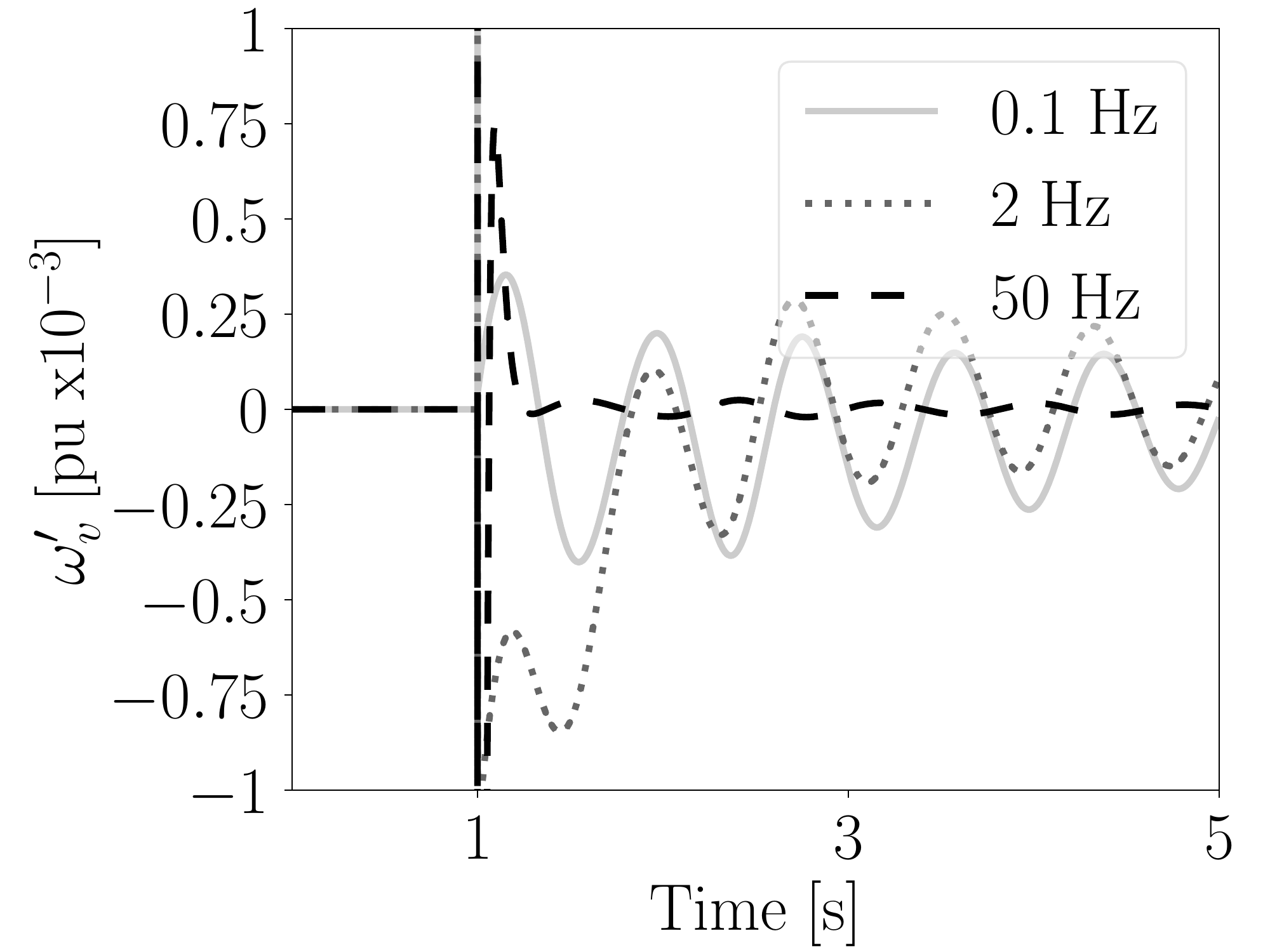}
\includegraphics[width=0.49\columnwidth]{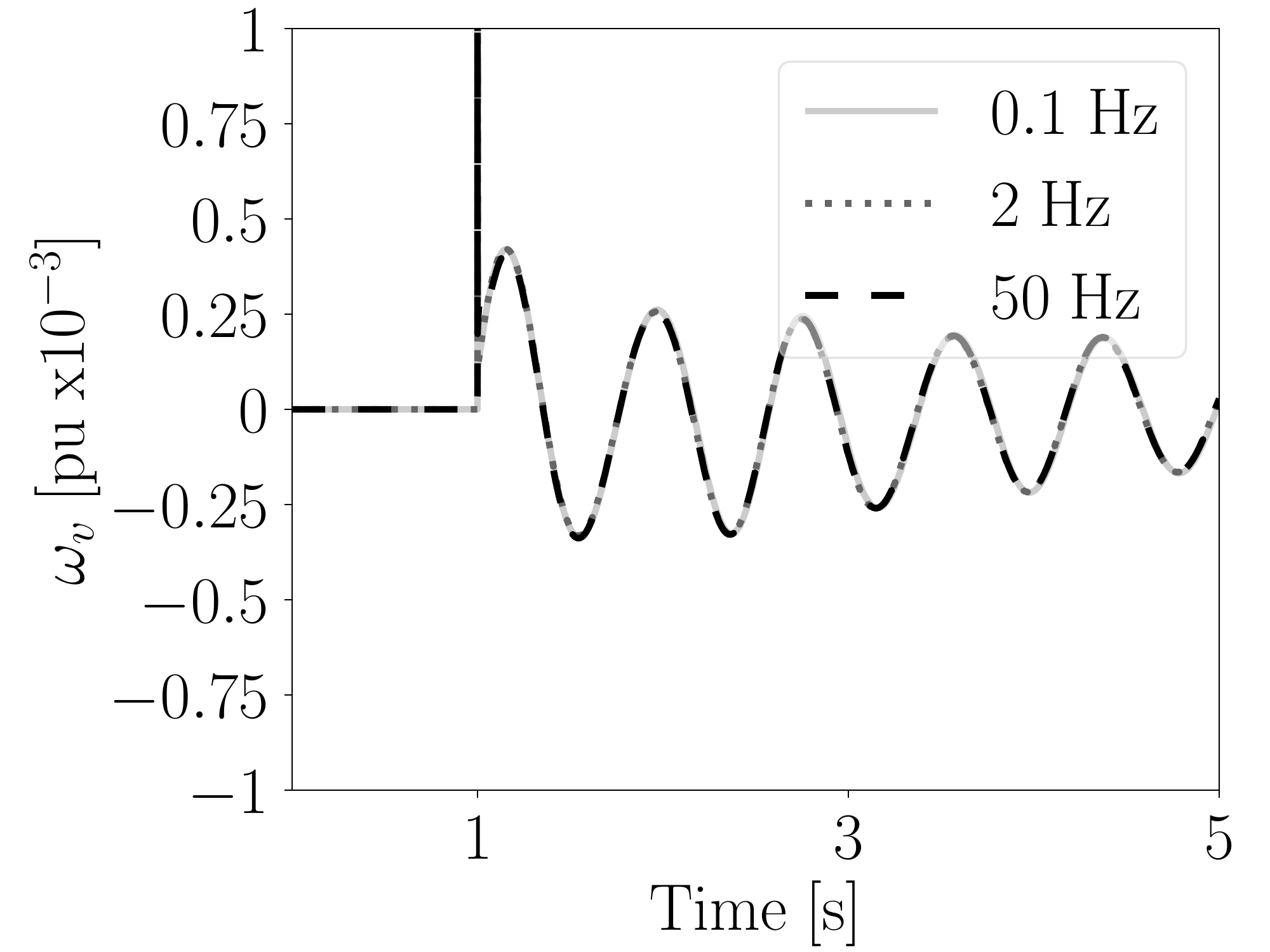}
\caption{Frequency deviation referred to the converter reference frame and as seen at bus 5 of the WSCC 9-bus system during the disconnection at $t=1$~s of the load at bus 5. \ac{pll} gains are set for various controller bandwidth values.}
\label{fig.wpll}
\vspace{-2mm}
\end{figure}
\begin{figure}
\centering
\includegraphics[width=0.49\columnwidth]{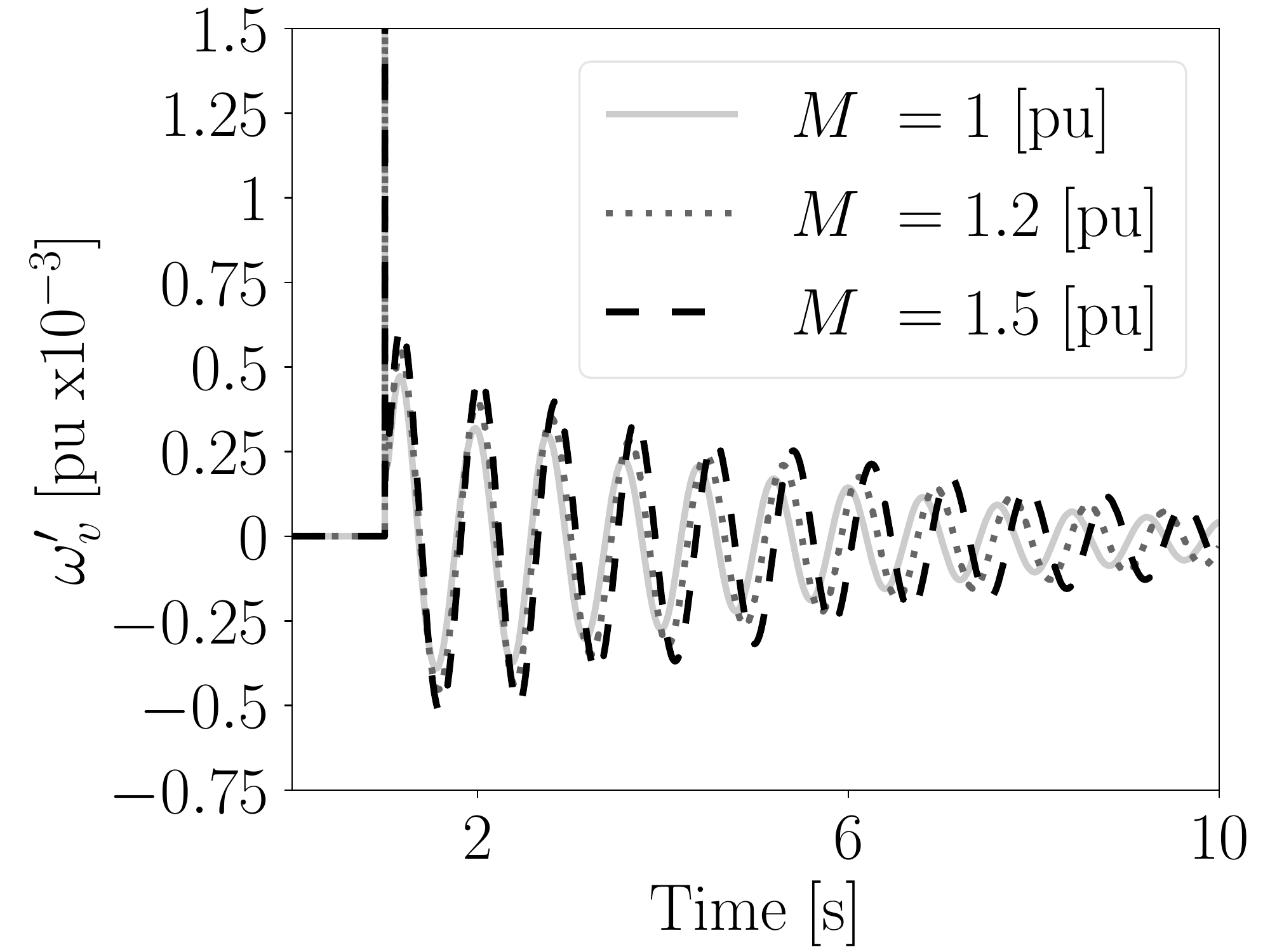}
\includegraphics[width=0.49\columnwidth]{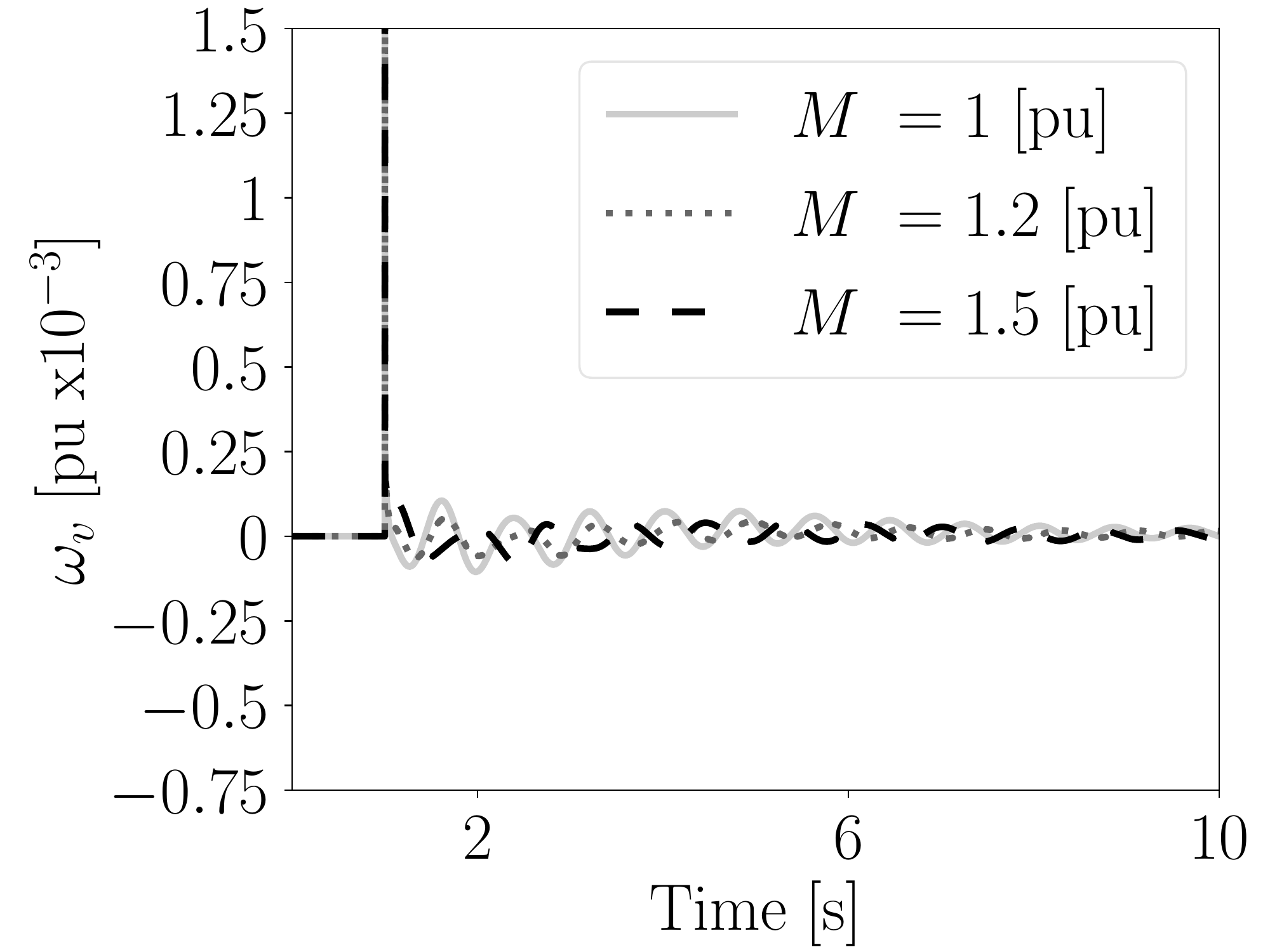}
\caption{Frequency deviation referred to the synchronous machine rotor and as seen at bus 1 of the WSCC 9-bus system during the disconnection at $t=1$~s of the load at bus 5. The simulation is repeated for different values of $M$.}
\label{fig.wsync}
\vspace{-2mm}
\end{figure}
Fig.~\ref{fig.wpll} shows the imaginary part of the \ac{cf}
$\omega_v'$ referred to the converter reference frame~(equation
\eqref{eq:vdot2}) as well as the frequency $\omega_v$, as seen at the
converter bus.  The \ac{pll} gains are tuned so that the \ac{pll}
damping is always one while its bandwidth varied.  While the frequency
at the bus is unaffected by the \ac{pll} gains, the frequency referred
to the converter is dictated by the \ac{pll} dynamics.
A parallel for \ac{pll} case can be drawn with the case of the
synchronous machine.  For that device, ``internal'' frequency
$\omega_v'$ can be calculated with equation~\eqref{eq:vdot2} where in
that case, $\dot{\delta}$ is the rotor speed referred to the frequency
at the center of inertia, namely
$\dot{\delta}=\omega_r-\omega_{\coi}$.
Fig.~\ref{fig.wsync} shows the frequency deviation referred to the
synchronous machine rotor and at bus 1 after the contingency for
different values of the mechanical starting time $M$.
The mechanical starting time is defined as $M=2 \; H$, where $H$ is
the inertia constant of a synchronous machine~\cite{milano2010power}.
The two quantities are often used interchangeably in the literature to
describe the inertia properties of a synchronous
machine~\cite{milano2010power, kundur2022power}.
The parameter change affects both quantities, but it can be seen that the internal frequency undergoes larger variations.
The above examples illustrates that while the internal frequency for the synchronous machine is well defined and understood, this is not the case for the different converter controllers.
The proposed method helps to overcome this problem by clearly defining it for the different synchronization methods.
It is shown in Section \ref{sub:ctrlapp} how this method can be used for control applications.
\subsection{Non-Ideality of the Current Controller}
\label{sec:casecc}
\begin{figure}
\centering
\includegraphics[width=0.49\columnwidth]{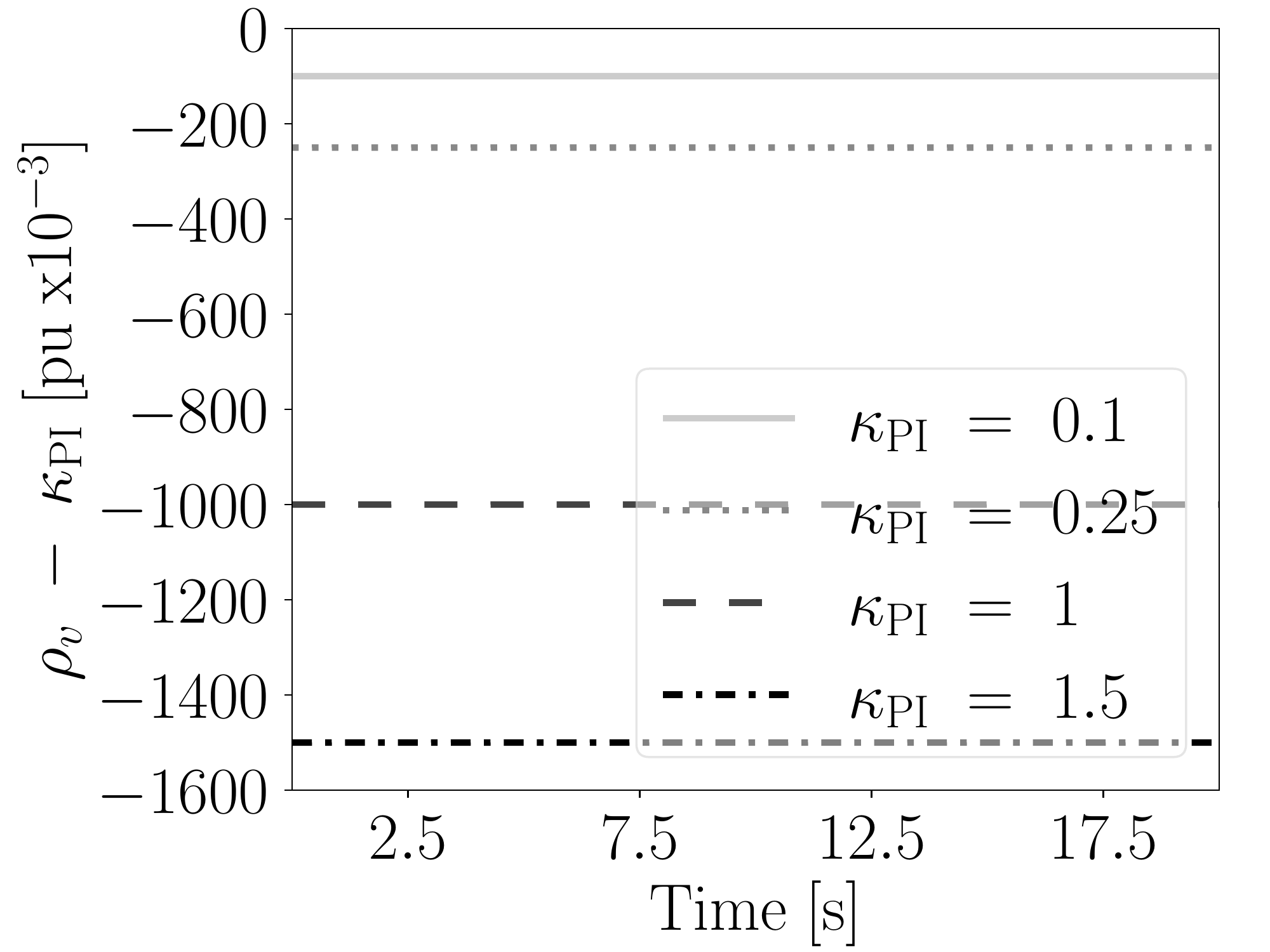}
\includegraphics[width=0.49\columnwidth]{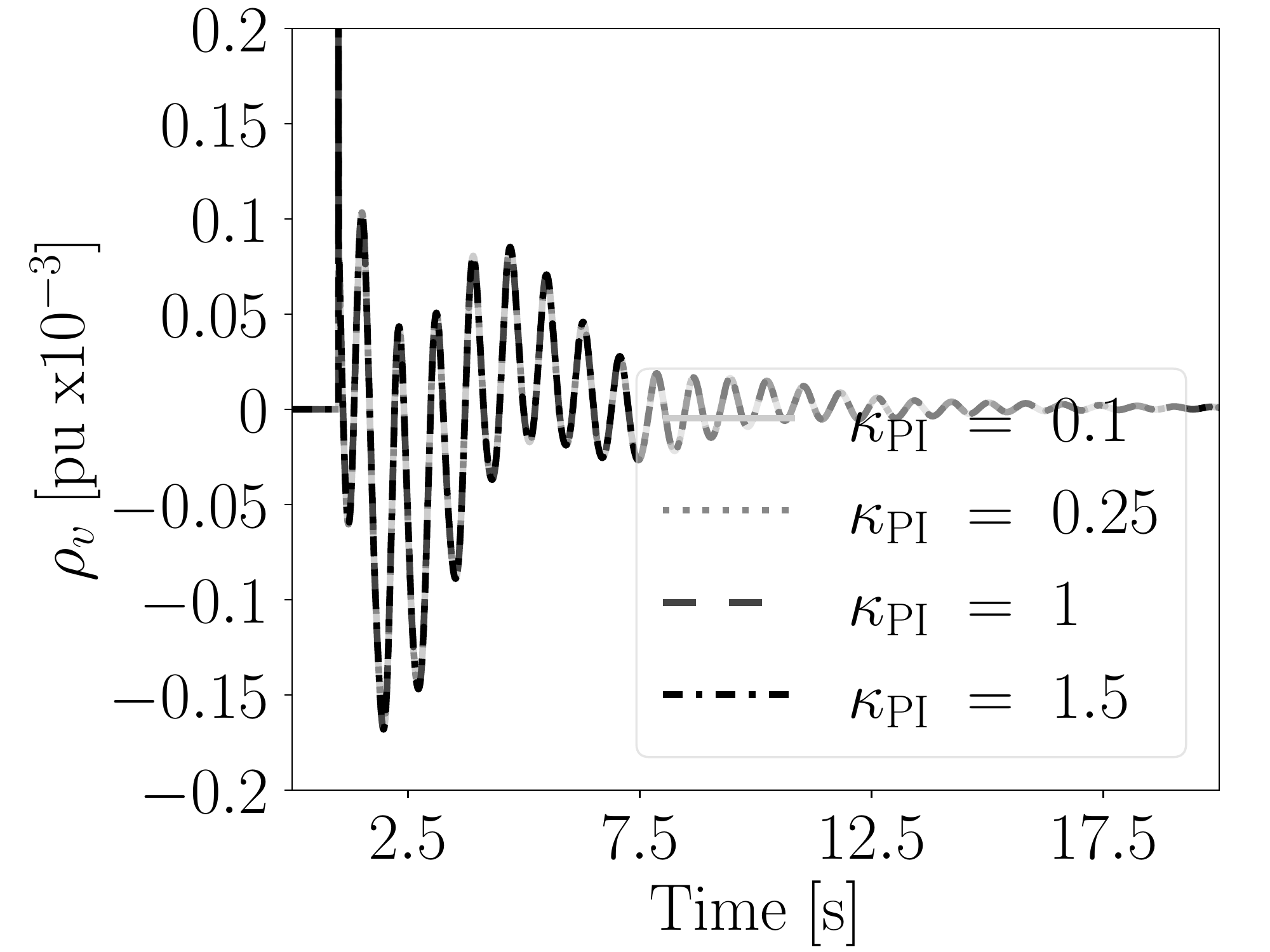}
\caption{Real part of the \ac{cf} $\rho_v$, translated by the effect of the PI current controller and as seen at bus 5 during the contingency for different values of $\kpi{}$.}
\label{fig.rhocc}
\vspace{-2mm}
\end{figure}
\begin{figure}
\centering
\includegraphics[width=0.49\columnwidth]{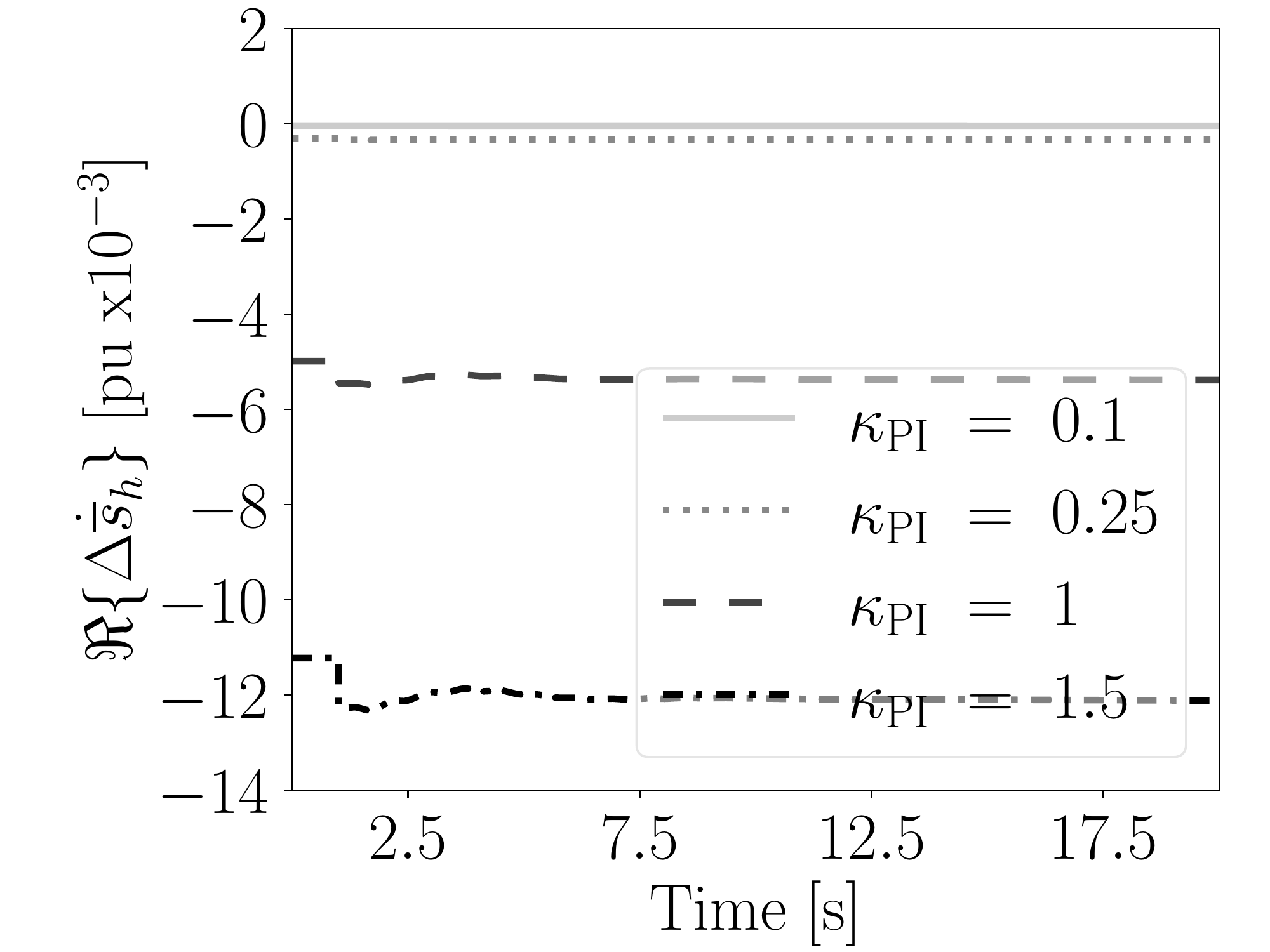}
\includegraphics[width=0.49\columnwidth]{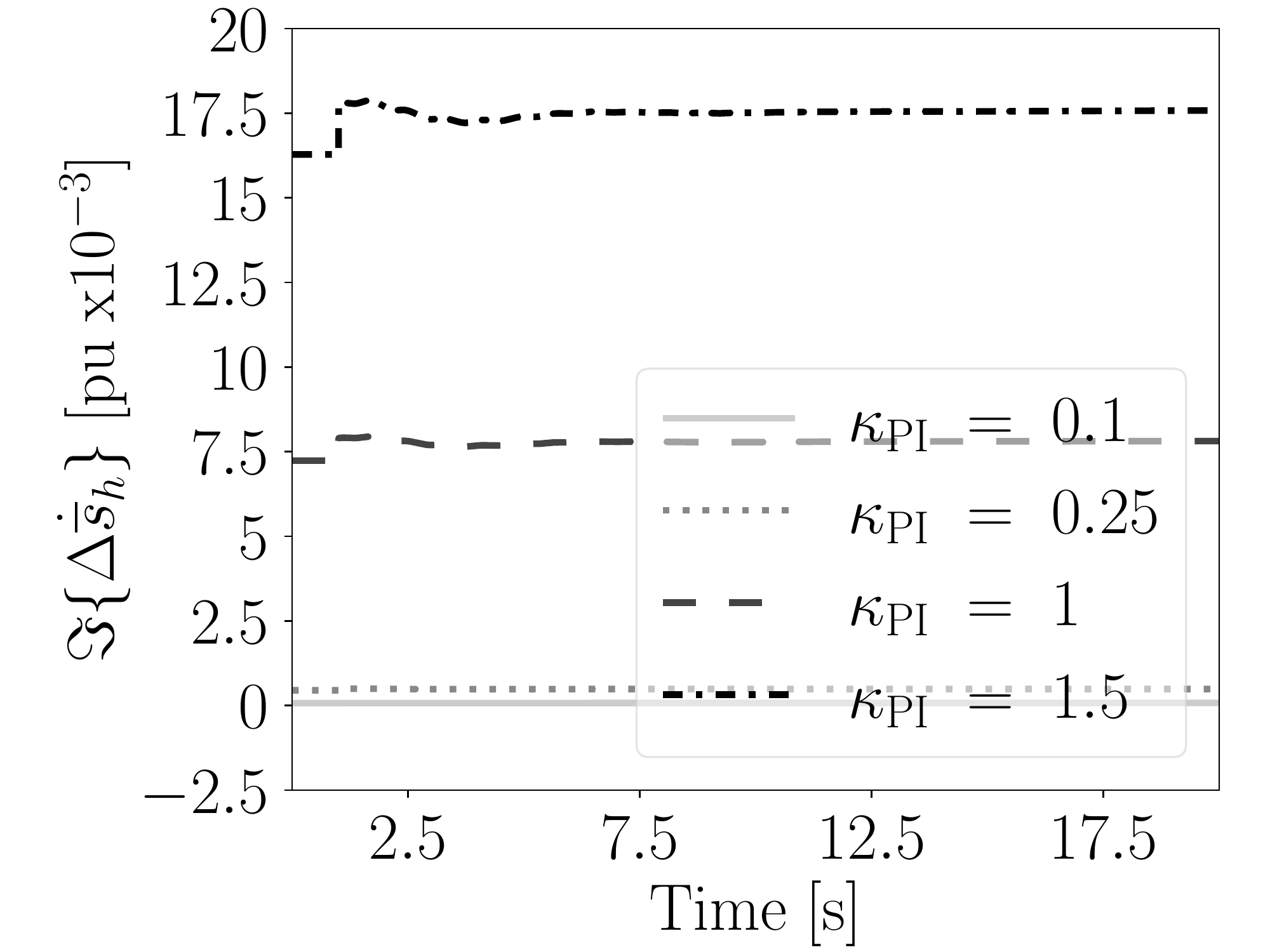}
\caption{Real and imaginary part of the quantity $\Delta \cpd{s}_h$, quantifying the non-ideality of the current control. Simulation for disconnection at $t=1$~s of the load at bus 5 and different values of $\kpi{}$.}
\label{fig.deltasdot}
\vspace{-2mm}
\end{figure}
Fig.~\ref{fig.rhocc} shows the translation in the \ac{cf} caused by the internal current PI as well as the real part of the \ac{cf} at the converter bus.
It can be seen that the variation of $\rho_v$ is negligible compared to the constant translation caused by the PI.
A better way to illustrate the non-ideality of the current controller is by plotting its deviation from an ideal current controller $\Delta \cpd{s}_h$, calculated with \eqref{eq:sdot_ccc2}.
Fig.~\ref{fig.deltasdot} shows the real and imaginary parts of this quantity for different values of parameter $\kpi{}$.
For larger values of the parameter, the operation of the controller increasingly deviates from the ideal case, namely $\kpi{}=0$.
\subsection{Effect of the \ac{vff}}
\begin{figure}
\centering
\includegraphics[width=0.825\columnwidth]{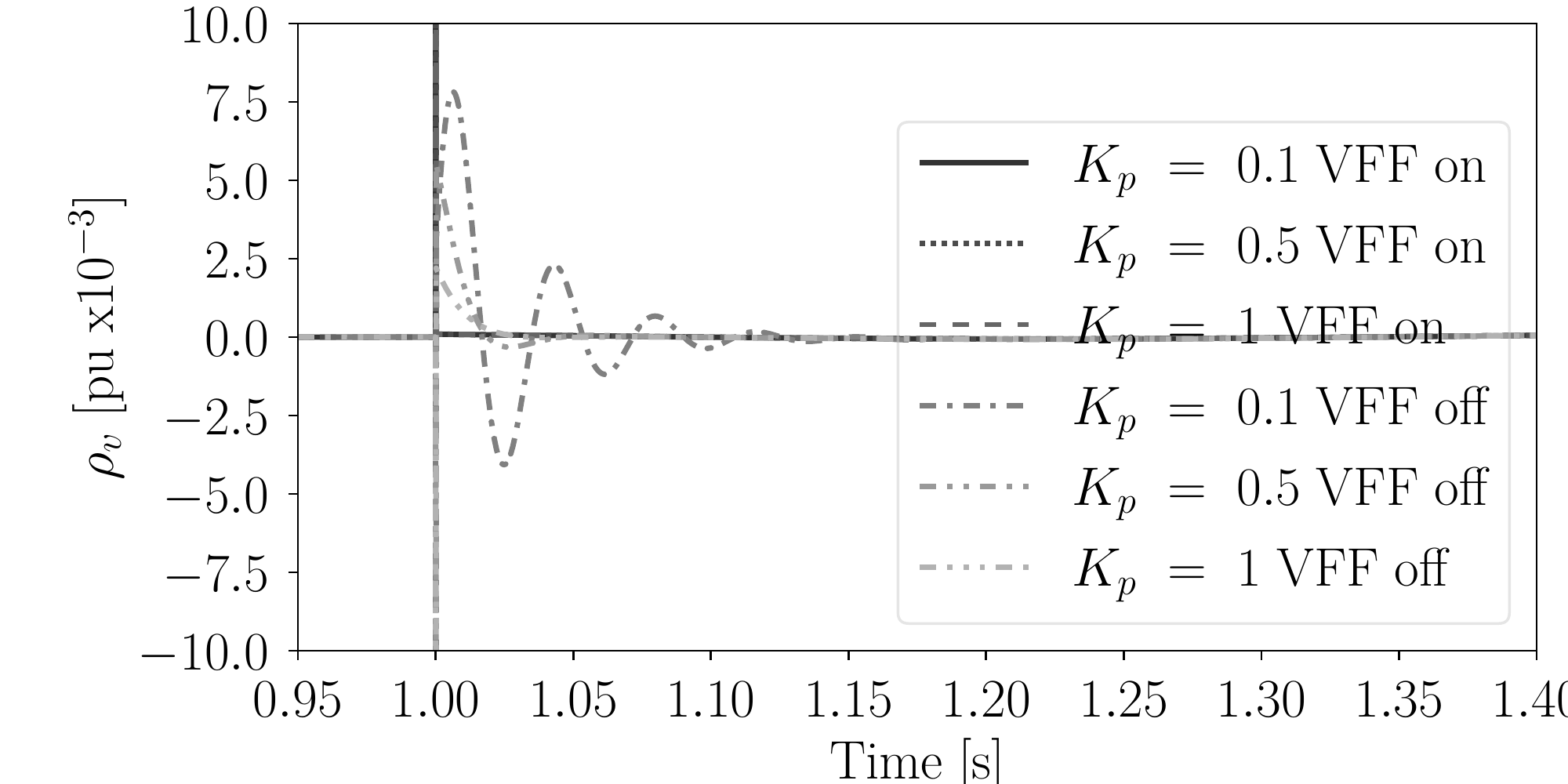}
\caption{Real part of the \ac{cf} as seen at bus 5 of the WSCC 9-bus system during the disconnection at $t=1$~s of the load at bus 5. Simulation for different values of parameter $K_p$ and \ac{vff} enabled and disabled.}
\label{fig.rhovff}
\vspace{-2mm}
\end{figure}
\begin{figure}
\centering
\includegraphics[width=0.825\columnwidth]{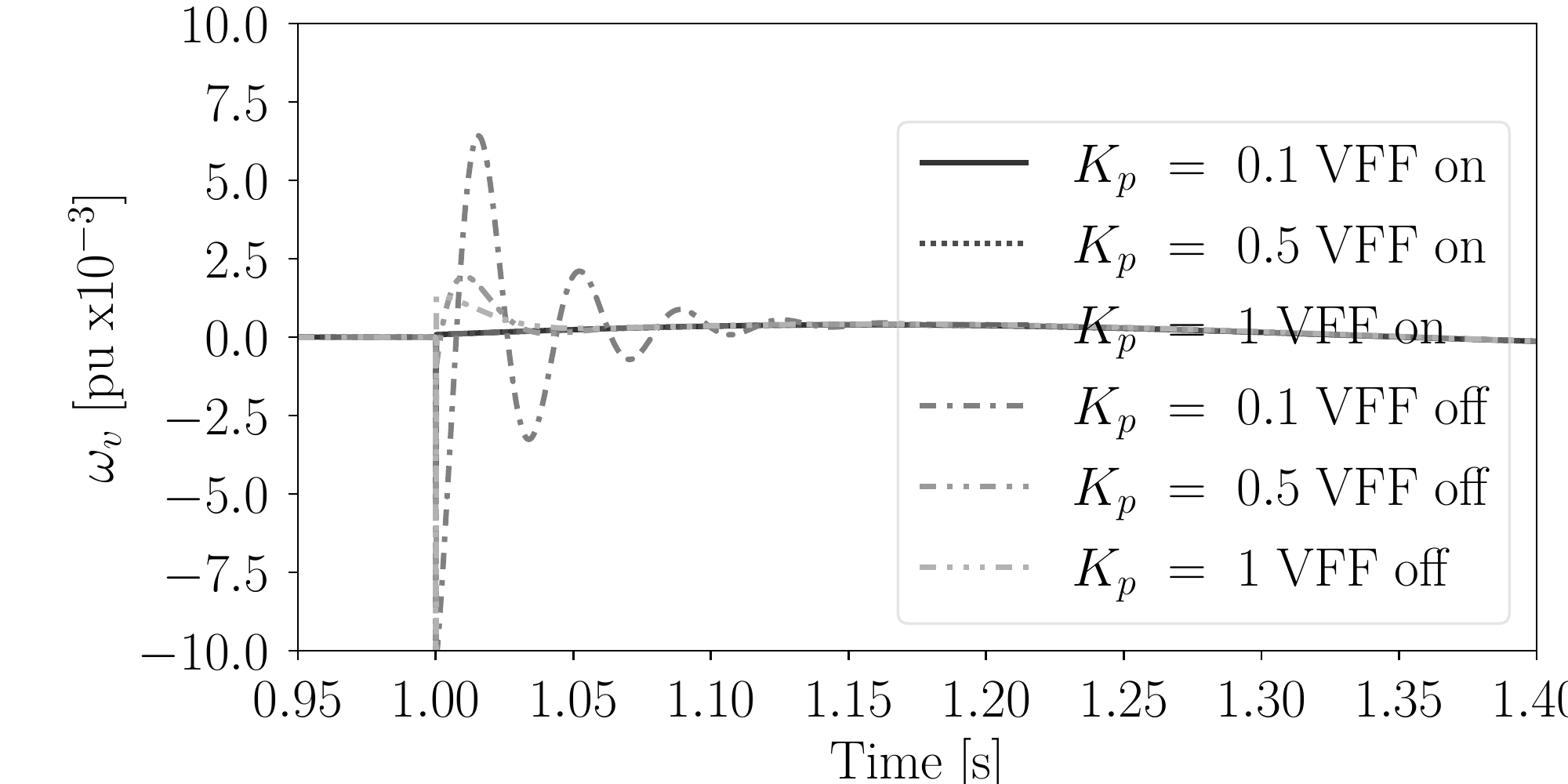}
\caption{Imaginary part of the \ac{cf} as seen at bus 5 of the WSCC 9-bus system during the disconnection at $t=1$~s of the load at bus 5. Simulation for different values of parameter $K_p$ and \ac{vff} on and off.}
\label{fig.omegavff}
\vspace{-2mm}
\end{figure}
Figs. \ref{fig.rhovff} and~\ref{fig.omegavff} show the real and imaginary part of the \ac{cf} at bus 5 for different values of current control proportional gain $K_p$.
The simulation is repeated with the \ac{vff} being switched on and off for each value of the controller gain.
If the \ac{vff} is active, the change of the parameter does not affect neither of the two components of the \ac{cf}.
If the \ac{vff} is switched off, the gain modulates the frequency, as predicted by \eqref{eq:sdotvff2}.
\subsection{Effect of the Outer Loops}
\begin{figure}
\centering
\includegraphics[width=0.49\columnwidth]{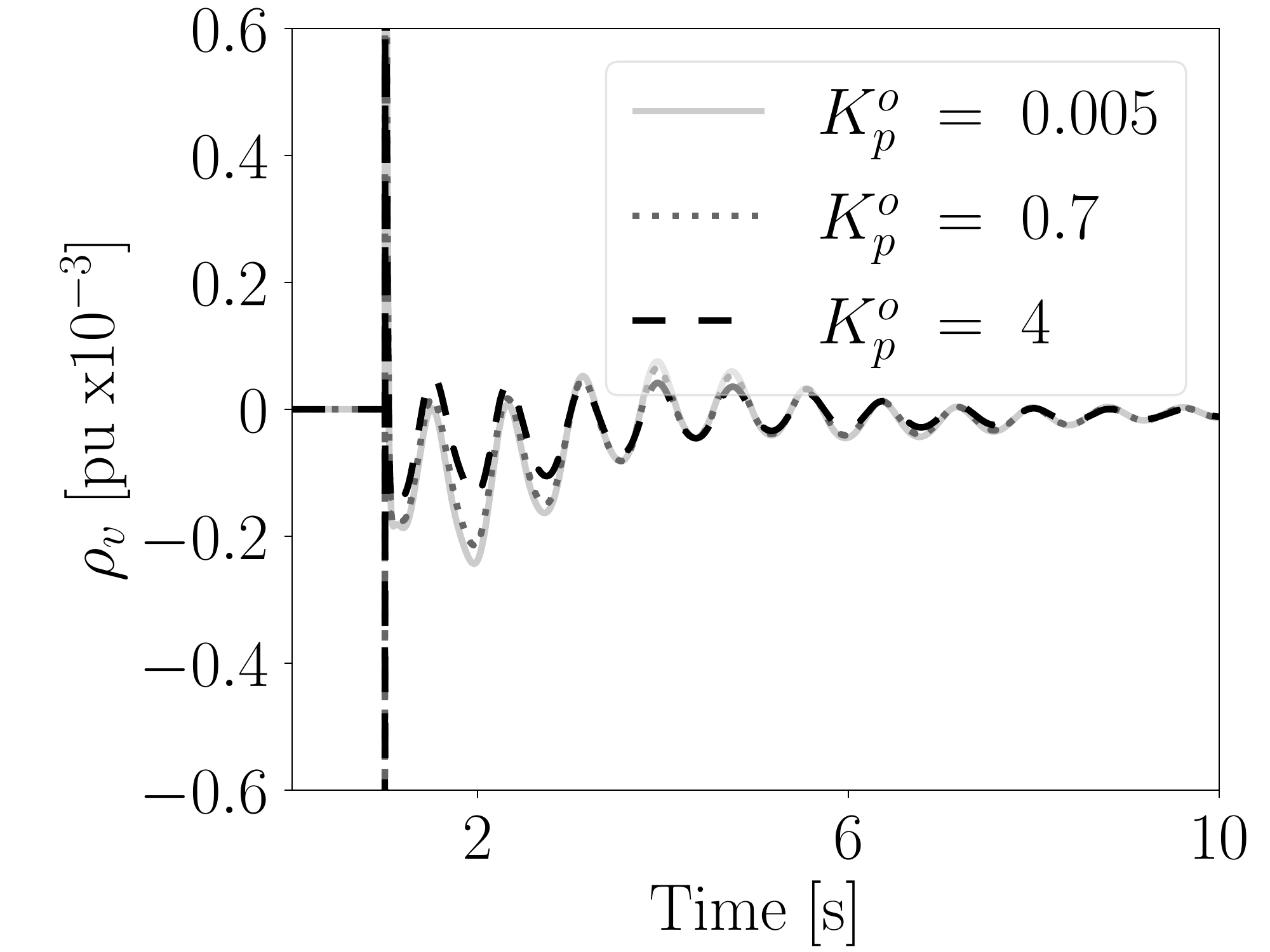}
\includegraphics[width=0.49\columnwidth]{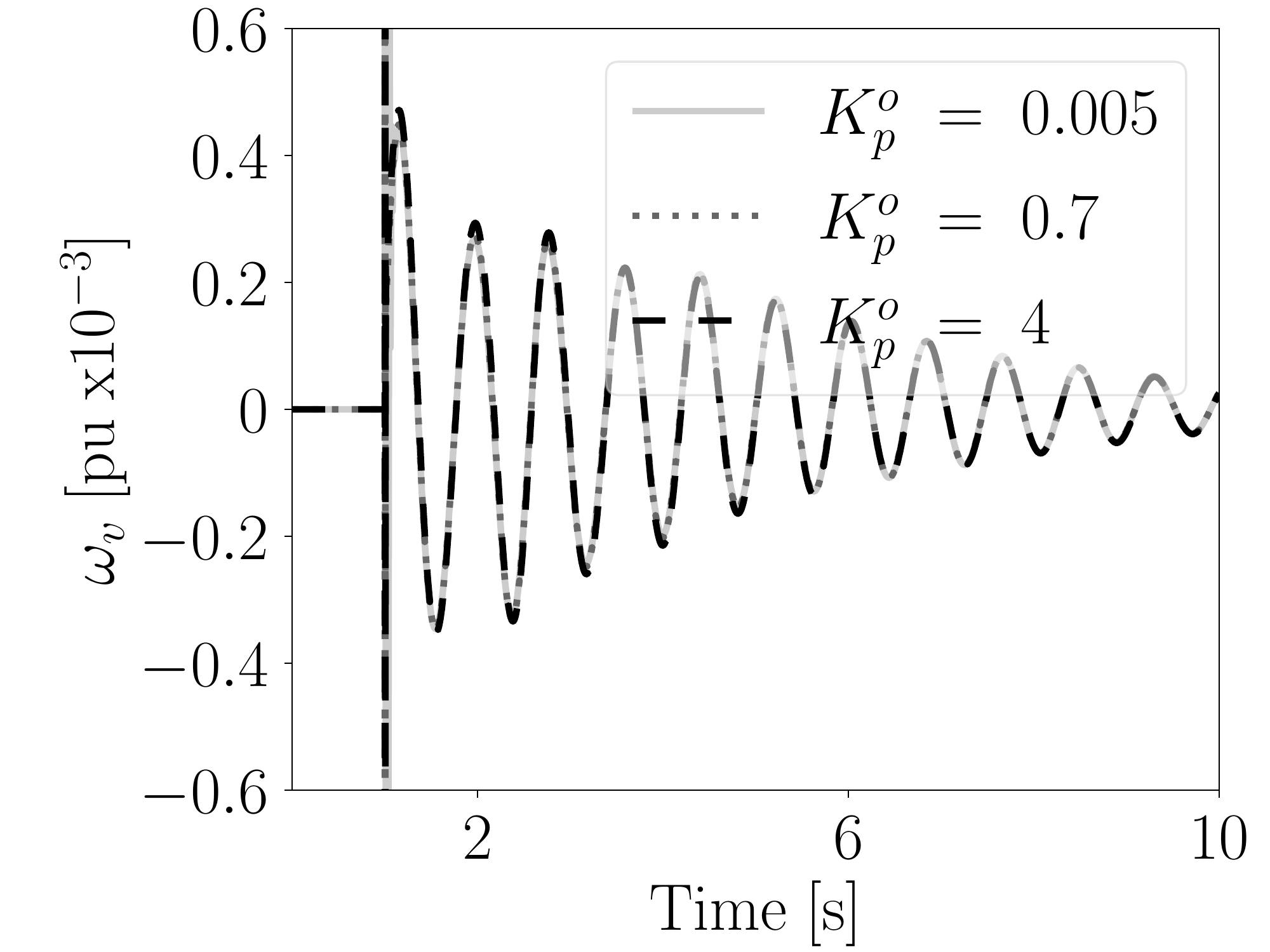}
\caption{Real and imaginary part of the \ac{cf} as seen at bus 5 of the WSCC 9-bus system during the disconnection at $t=1$~s of the load at bus 5. Simulation for different values of parameter $K_p^o$ of the outer voltage loops in the \ac{gfl} control configuration.}
\label{fig.outerloops}
\end{figure}
\begin{figure}
\centering
\includegraphics[width=0.825\columnwidth]{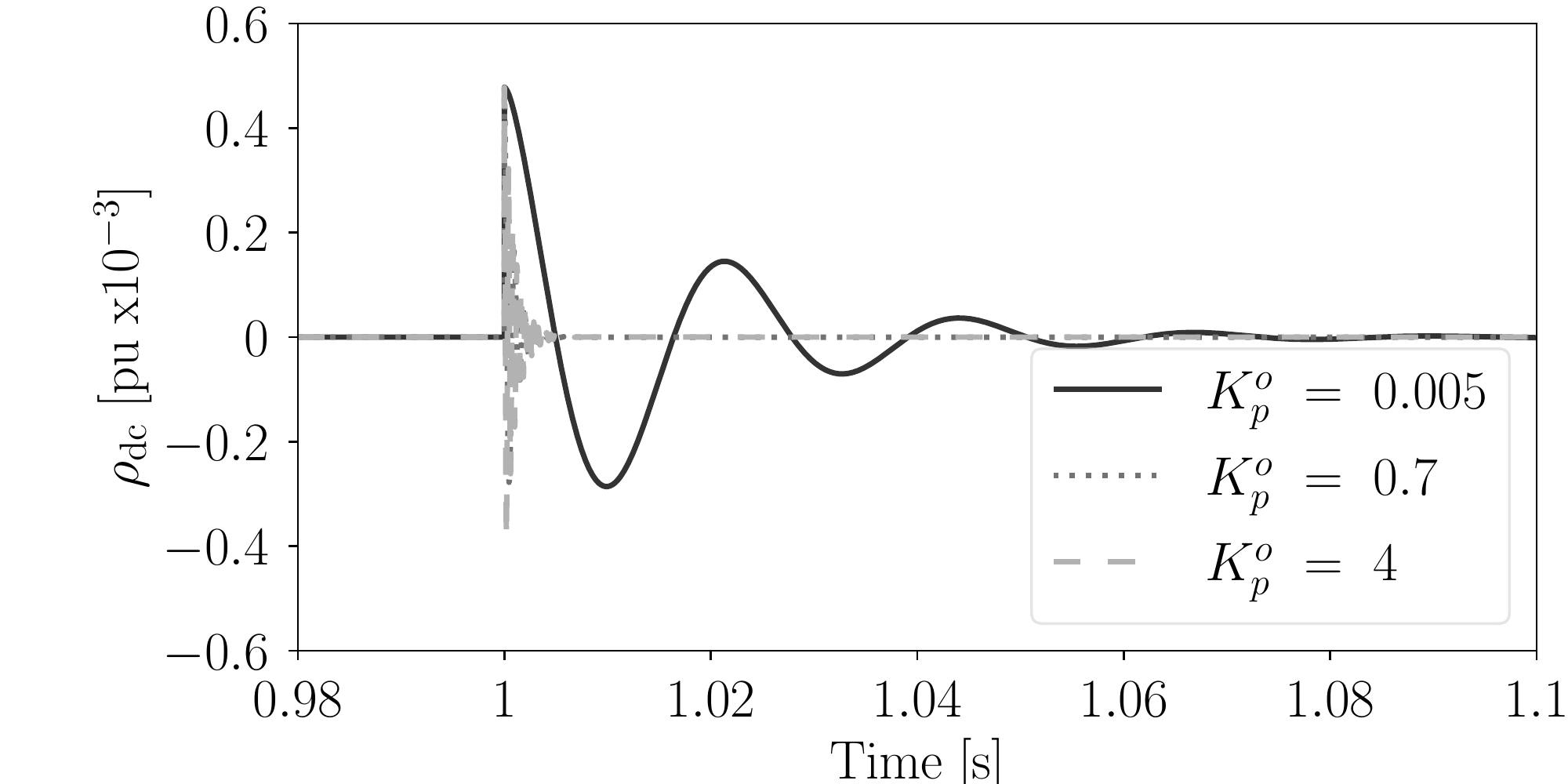}
\caption{dc ``frequency'' during the contingency at $t=1$~s. Simulation for different values of parameter $K_p^o$ of the outer voltage loops in the \ac{gfl} control configuration.}
\label{fig.rhodc}
\vspace{-2mm}
\end{figure}
Fig.~\ref{fig.outerloops} shows the effect on the \ac{cf} of the variation of gain $K_p^o$ of the outer control loops in the \ac{gfl} control of \eqref{eq:outer_loop_complex}.
Results confirm that only $\rho_v$ is affected while $\omega_v$ remains identical.
Fig.~\ref{fig.rhodc} shows the variation of $\rho_{\rm dc}$ for the same set of parameters.
For this specific dc-side topology, $\rho_{\rm dc}$ is calculated by standard circuit variables available from the software~($\rho_{\rm dc}=\dot{v}_{\rm dc}/v_{\rm dc}=\ii_{\rm dc}/(C_{\rm dc} v_{\rm dc}$)).
Although the transient effect diminishes quickly for higher gain values, it can be seen that changing gain $K_p^o$ has the expected effect on the quantity $\rho_{\rm dc}$.
The modulation of both $\rho_v$, $\rho_{\rm dc}$ by parameter $K_p^o$ justifies the use of the generalized quantity $\cp{\rho}$ in \eqref{eq:outer_loop_complex}.
The use of identical gains for the $dq$ channels, without loss of stability, also justifies the complex notation of \eqref{eq:outer_loop_complex}.
For this specific application, the ratio of dc to ac voltage is: $v_{n, \rm dc}/v_n=0.35$.
\subsection{Effect of the Active Power Droop}
\begin{figure}
\centering
\includegraphics[width=0.49\columnwidth]{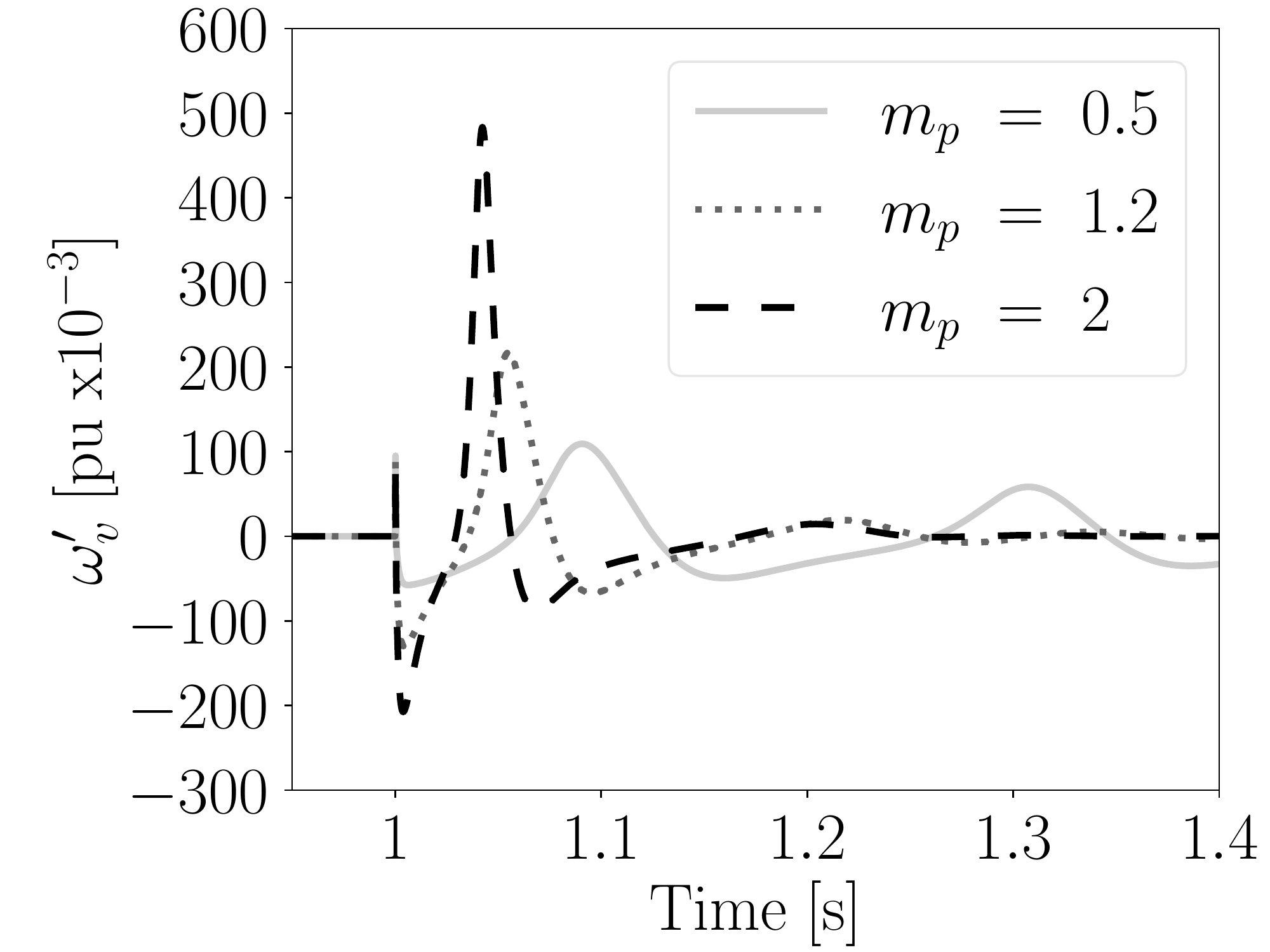}
\includegraphics[width=0.49\columnwidth]{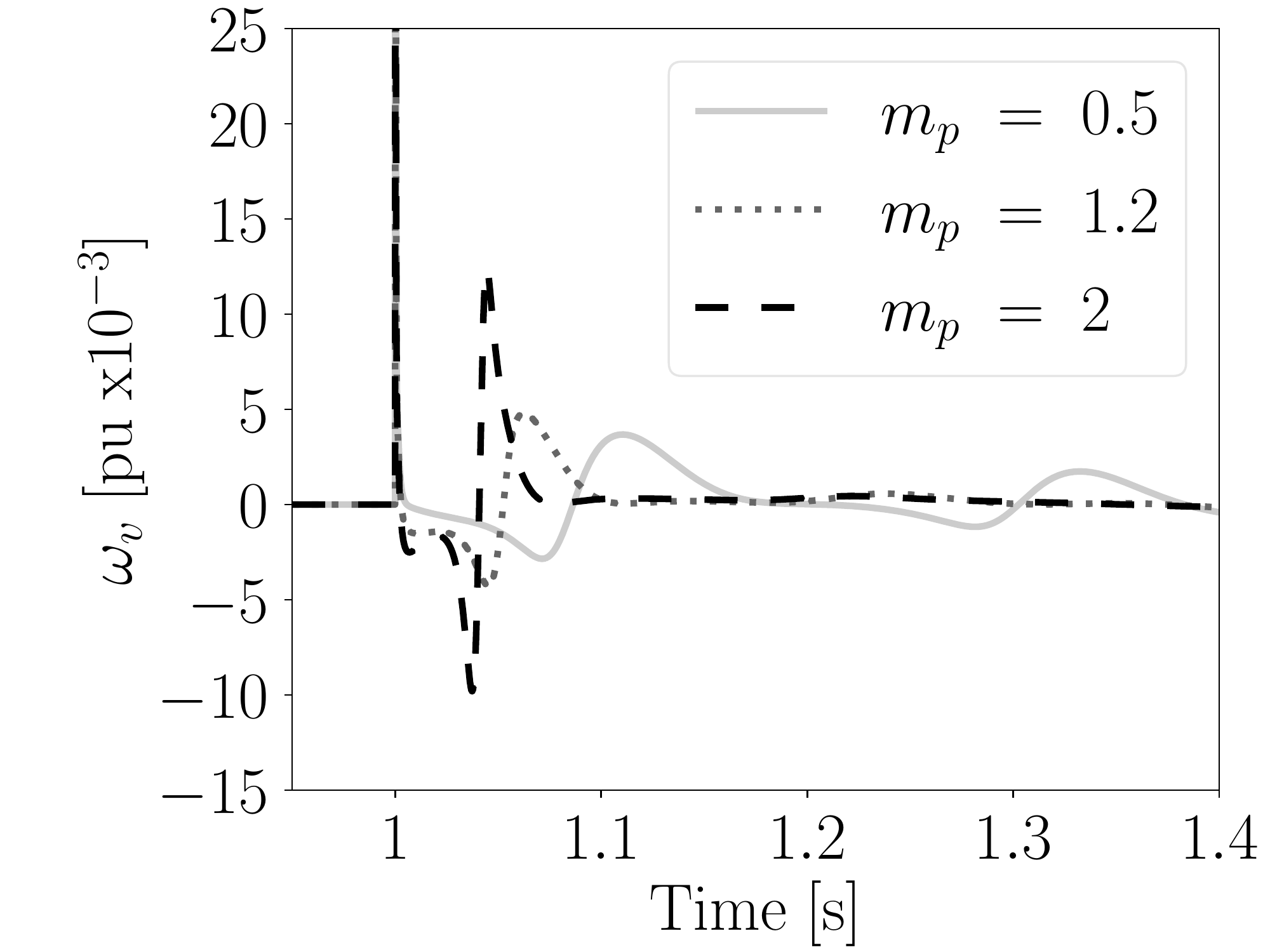}
\caption{Frequency deviation referred to the converter reference frame and as seen at bus 5 of the WSCC 9-bus system during the disconnection at $t=1$~s of the load at bus 5. The simulation is repeated for different values of the active power droop parameter $m_p$.}
\label{fig.droop}
\vspace{-2mm}
\end{figure}
Fig.~\ref{fig.droop} shows the effect of the active power droop parameter $m_p$ on the internal frequency and on the frequency at the converter bus after the contingency.
The effect of this outer-loop parameter on the frequency is more impactful than the previous cases that concern controllers with faster time scales.
Particularly for the internal frequency, the time response is separated by a full order of magnitude compared to the frequency at the bus.
Finally, note that although the effects are larger for this case, the selected unit for the figures~(pu x$10^{-3}$) is kept the same with the previous examples for consistency.
\subsection{Effect of the Virtual Shaft}
\begin{figure}
\centering
\includegraphics[width=0.49\columnwidth]{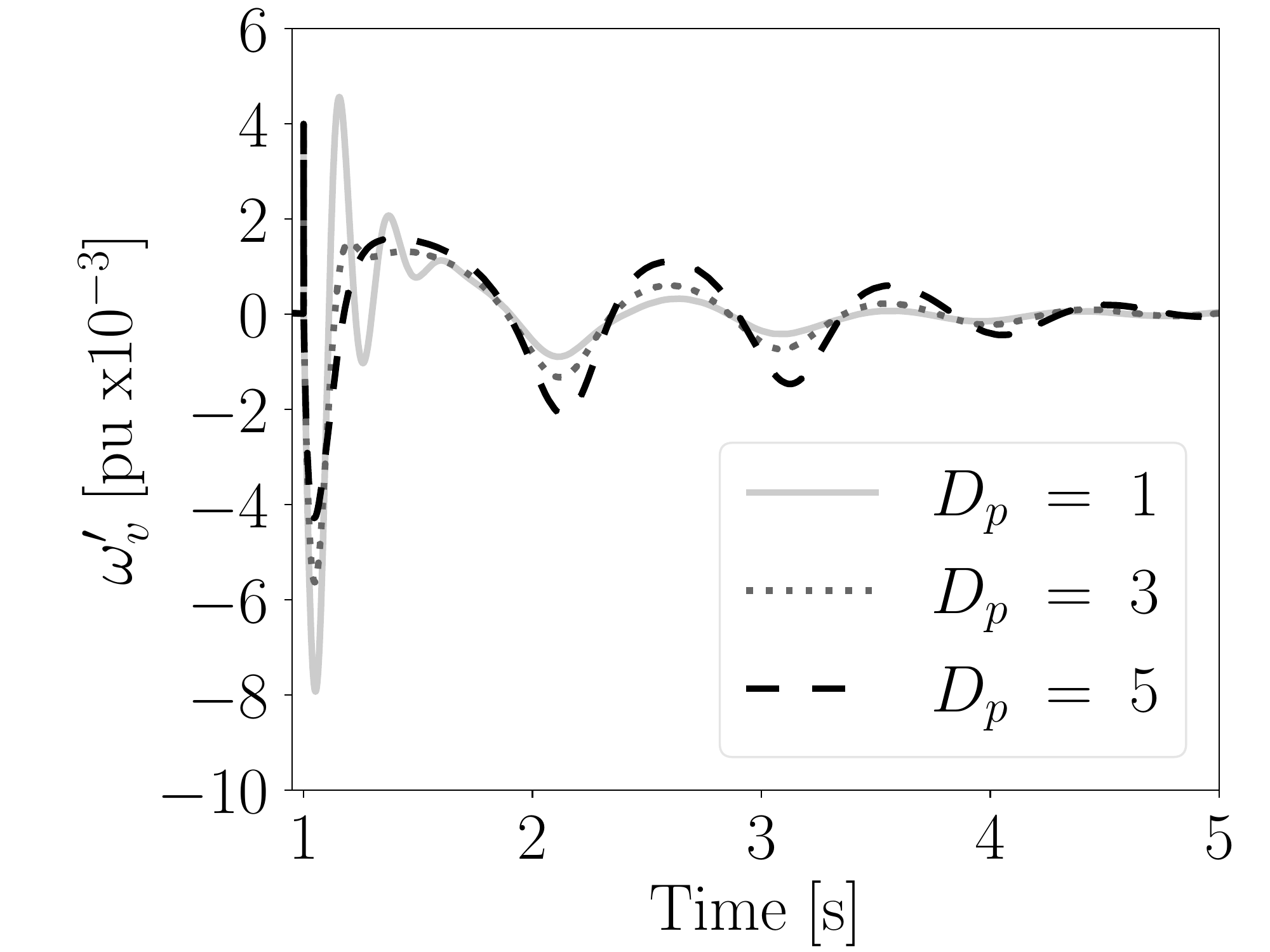}
\includegraphics[width=0.49\columnwidth]{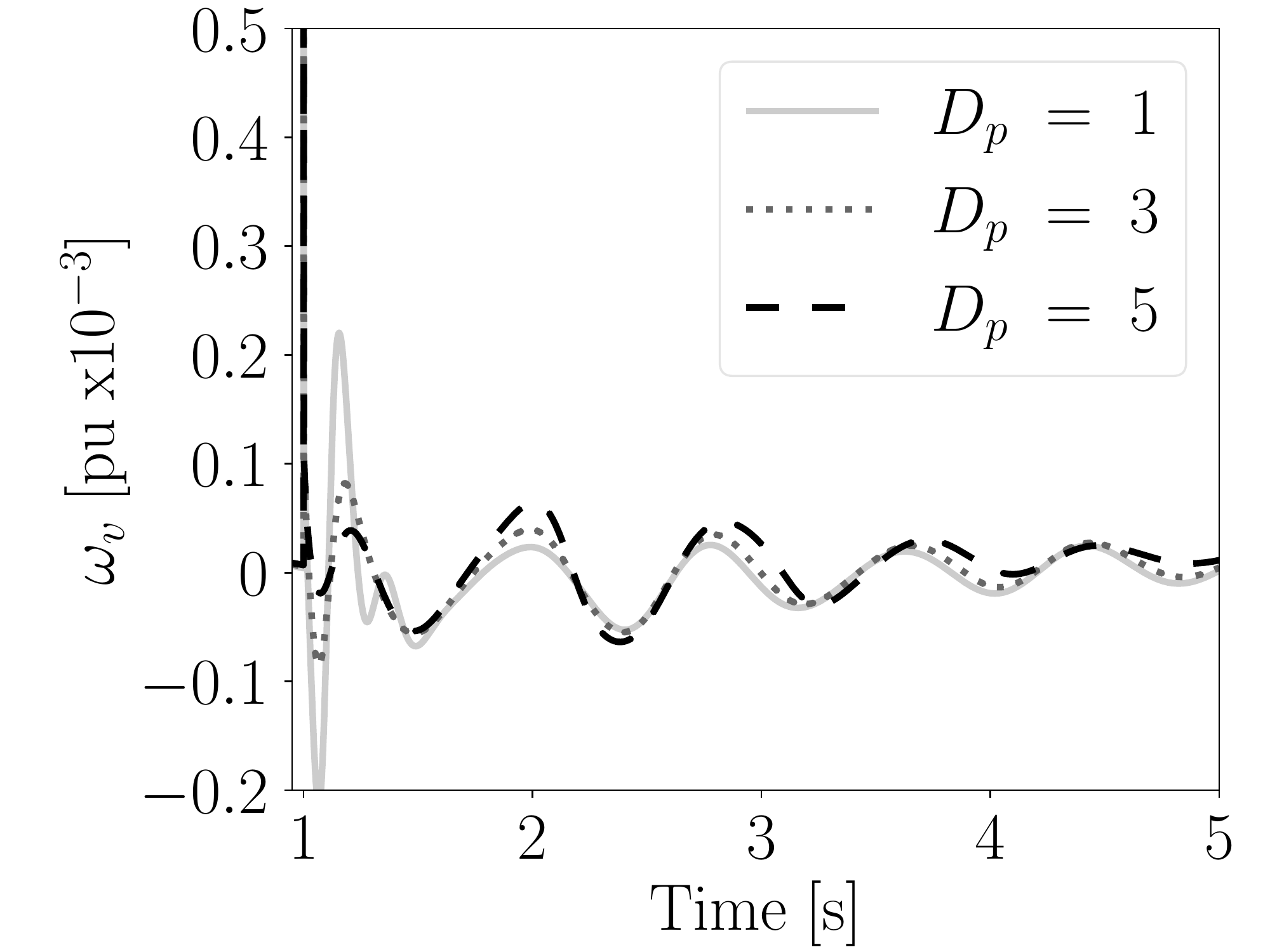}
\caption{Frequency deviation referred to the converter reference frame and as seen at bus 5 of the WSCC 9-bus system during the disconnection at $t=1$~s of the load at bus 6. The simulation is repeated for different values of the \ac{vsm} damping parameter $D_p$.}
\label{fig.vsm}
\vspace{-2mm}
\end{figure}
Fig.~\ref{fig.vsm} shows the frequency response when the damping parameter $D_p$ of the \ac{vsm} control of \eqref{eq:vsm} is changed.
The contingency for this case is a disconnection of the load at bus 6 at $t=1$~s.
Similar to the previous cases, the variation of the control parameters affects the internal frequency of the converter to a greater extent than the one at the bus. 
The examples discussed in the next section show how this property can be utilized to improve the control of the converters.
\subsection{Control Applications}
\label{sub:ctrlapp}
This section illustrates the use of the converter \textit{internal frequency} on control applications by means of two examples, one for \ac{gfl} and one for \ac{gfm}. 
While the frequency at the bus of the converter can be estimated
(e.g., through a \ac{pll}), it is unclear whether it is the best
signal one can use.
For both \ac{gfl} and \ac{gfm} cases, a \ac{pfr} is added on top of
the outer control layers described in the previous sections.
Specifically, the \ac{pfr} is added on top of the power
controller~(equation \eqref{eq:power_reference} and
Fig.~\ref{fig.powerrefcalc}) for the \ac{gfl} case and on top of the
active power droop controller~(equation \eqref{eq:pf_droop} and
Fig.~\ref{fig.activesyncscheme}~(a)) for the \ac{gfm} case.
The \ac{pfr} is composed of a low-pass filter, a washout
filter and a hard-limit~\cite{sanniti2022curvature}.
The \ac{pfr} modifies the active power reference $\rf{p}$ by quantity
$\Delta \rf{p}$ based on an input frequency signal.
The signals that are used for the comparison are the imaginary part of
the \ac{cf} at the converter bus and the imaginary part of the
\textit{internal} converter frequency as give by equations
\eqref{eq:vdot2}, \eqref{eq:PLL} and \eqref{eq:pf_droop}.
The comparison of the dynamic performance of the primary controllers
based on the bus frequency and on the \textit{internal} ones is the
main objective of this section.

\begin{figure}
\centering
\includegraphics[width=0.825\columnwidth]{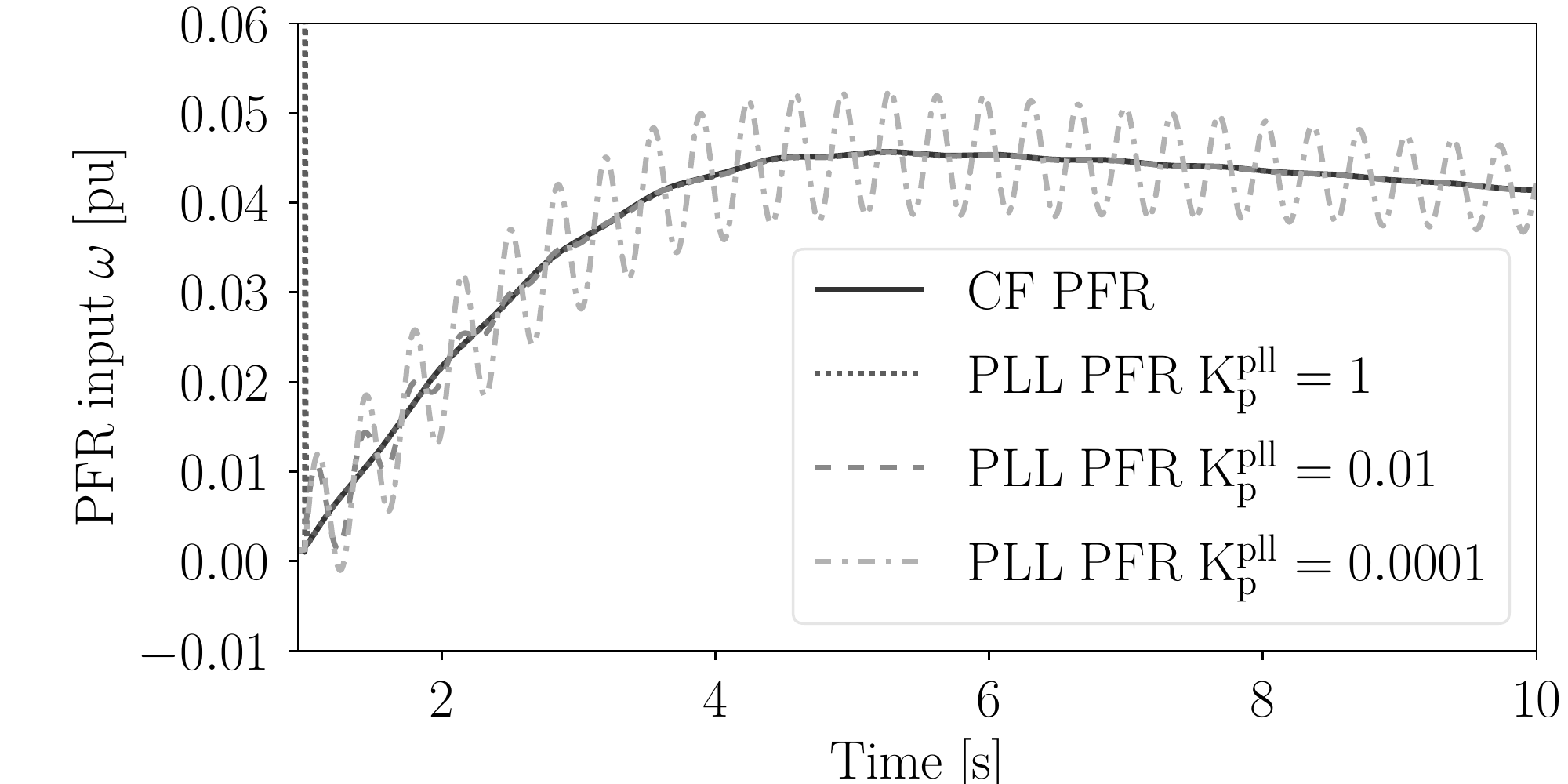}
\includegraphics[width=0.825\columnwidth]{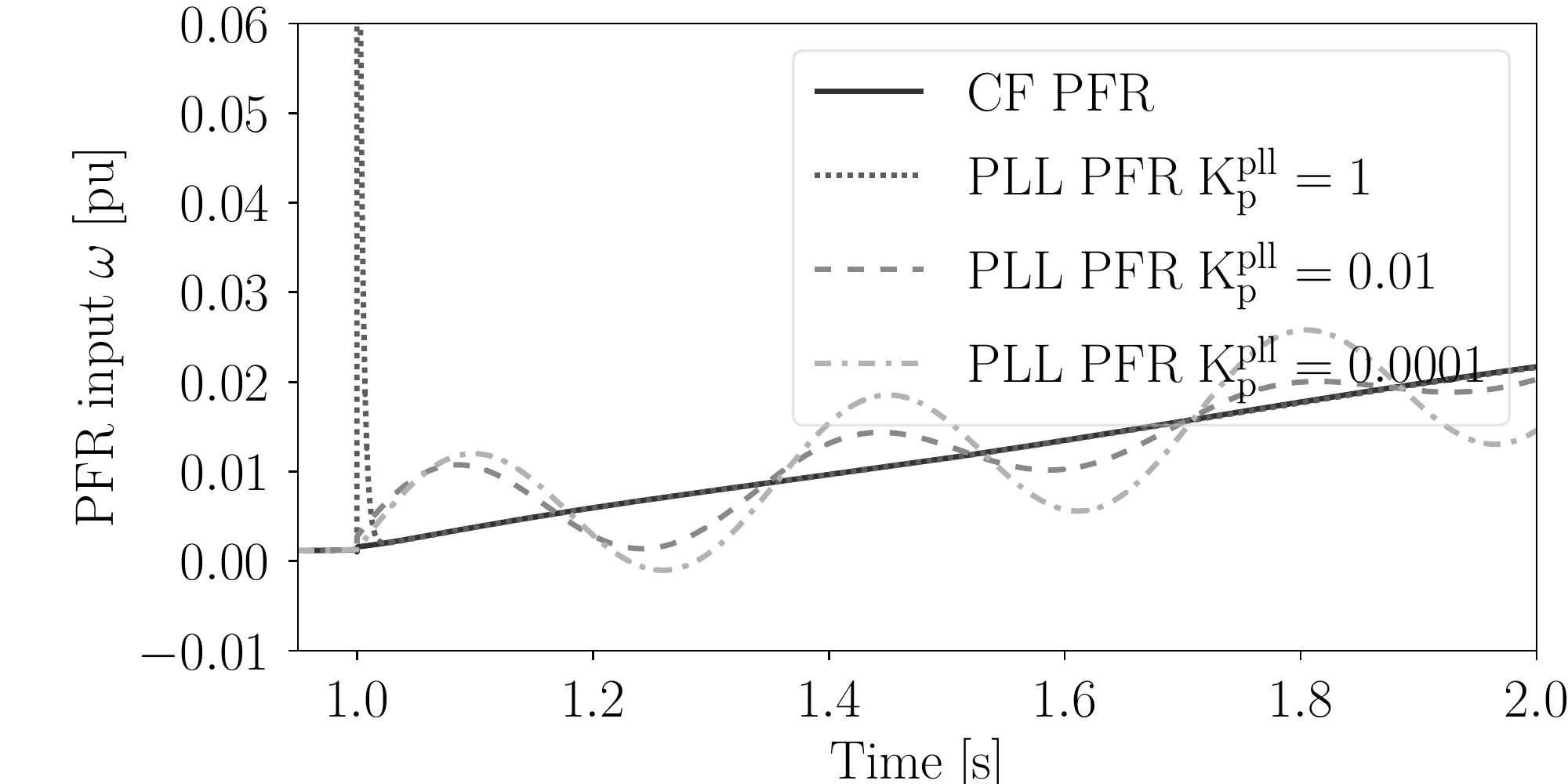}
\caption{Input signal (full and zoomed-in version) of the frequency
  control for the \ac{gfl} case. Disconnection at $t=1$~s of the load
  at bus 5 of the WSCC 9-bus system. The simulation is repeated for
  two types of input signal and different values of the \ac{pll}
  proportional gain $K_p^{\rm \pll}$.}
\label{fig.pfrinputgfl}
\end{figure}

\begin{figure}
\centering
\includegraphics[width=0.825\columnwidth]{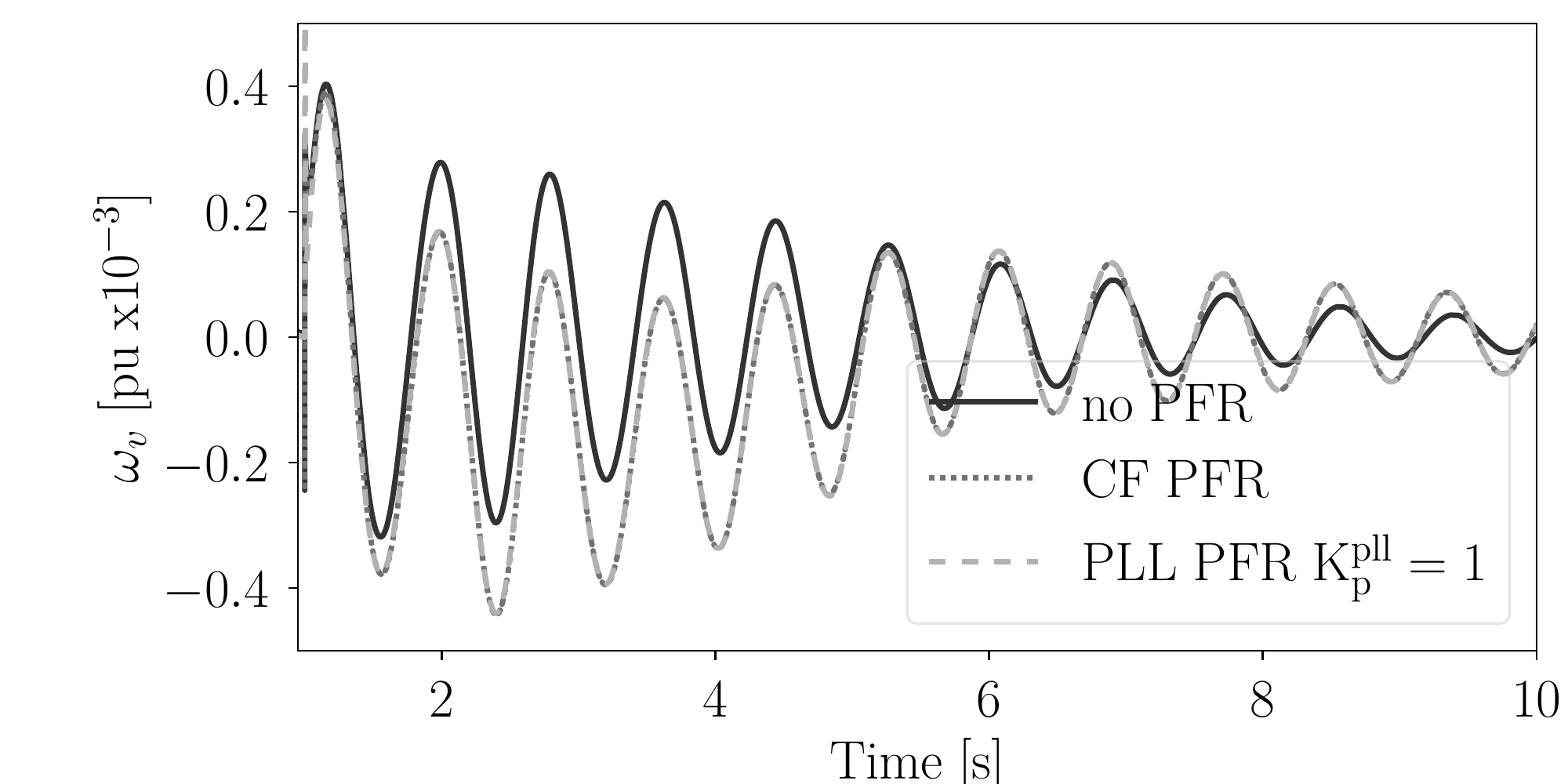}
\includegraphics[width=0.825\columnwidth]{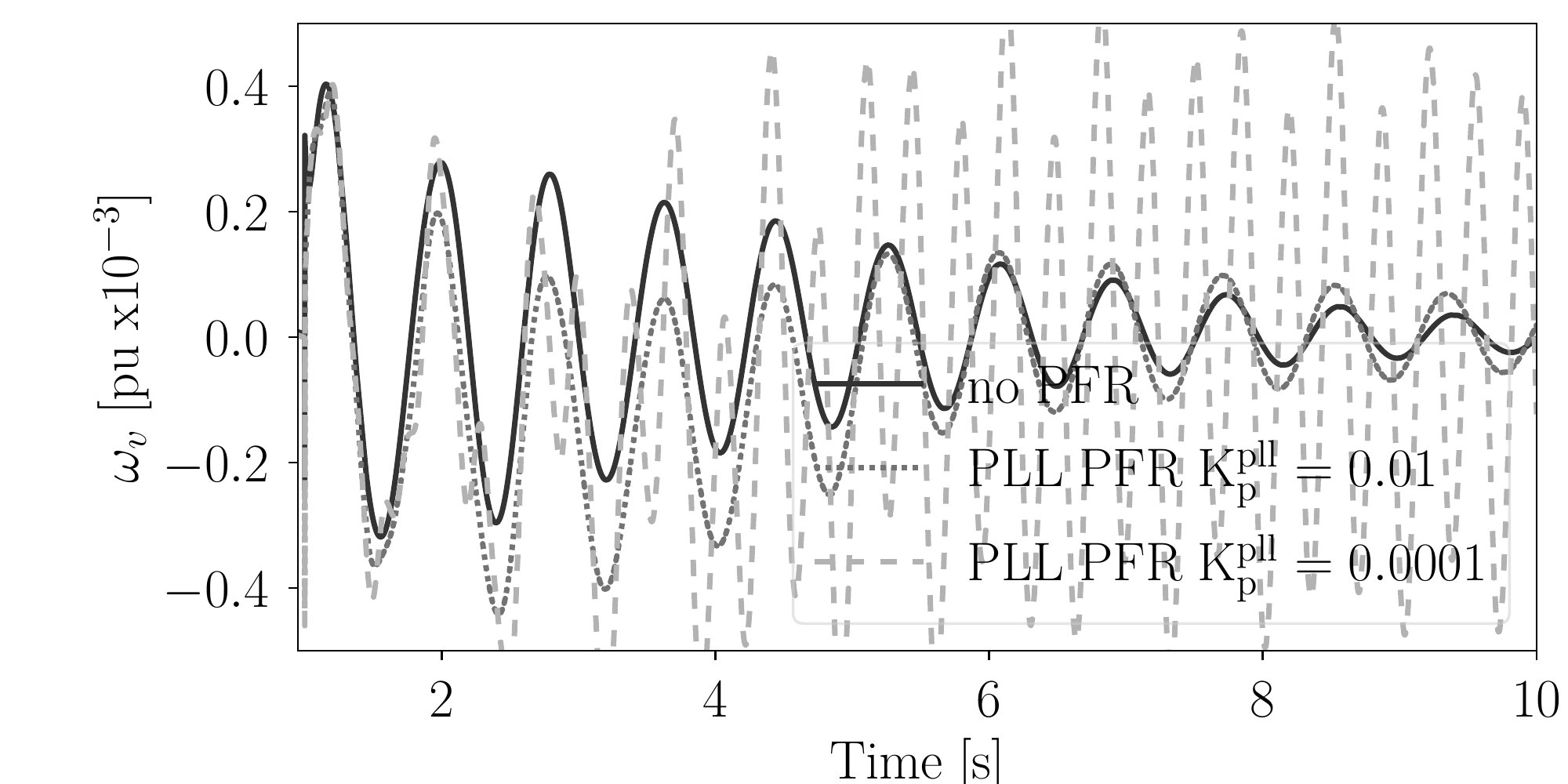}
\caption{Imaginary part of the \ac{cf} as seen at bus 5 of the WSCC 9-bus system during the disconnection at $t=1$~s of the load at bus 5. Application of frequency control for the \ac{gfl} case. The simulation is repeated without frequency control, for two types of input signal and for different values of the \ac{pll} proportional gain $K_p^{\pll}$.}
\label{fig.pfromegagfl}
\vspace{-2mm}
\end{figure}

Fig.~\ref{fig.pfrinputgfl} shows the \ac{pfr} input frequency signals
after the contingency for the \ac{gfl} case.
Different type of signals and different values of \ac{pll}
proportional gain $K_p^{\pll}$ are used.
In the zoomed-in version, it can be seen how the faster
\ac{pll}~($K_p^{\pll} = 1$) matches the exact frequency after the
initial transient.
The slower \acp{pll} introduce oscillations that remain for several seconds after the contingency.
The impact of this control on the bus frequency is shown in Fig.~\ref{fig.pfromegagfl}.
On the other hand, the faster the \acp{pll}, the better the frequency response.  This result is consistent with the case that uses the exact frequency signal.
It can be concluded that the internal frequency of the converter can have similar results with the exact frequency at the bus when used for frequency control.
This particularly useful for the cases where the exact signal is not available.
However, careful tuning of the \ac{pll} parameters should be considered so that its bandwidth is suitably fast for the application while preventing undesired coupling with the dynamics of (weak) grids \cite{7027822}.

\begin{figure}
\centering
\includegraphics[width=0.825\columnwidth]{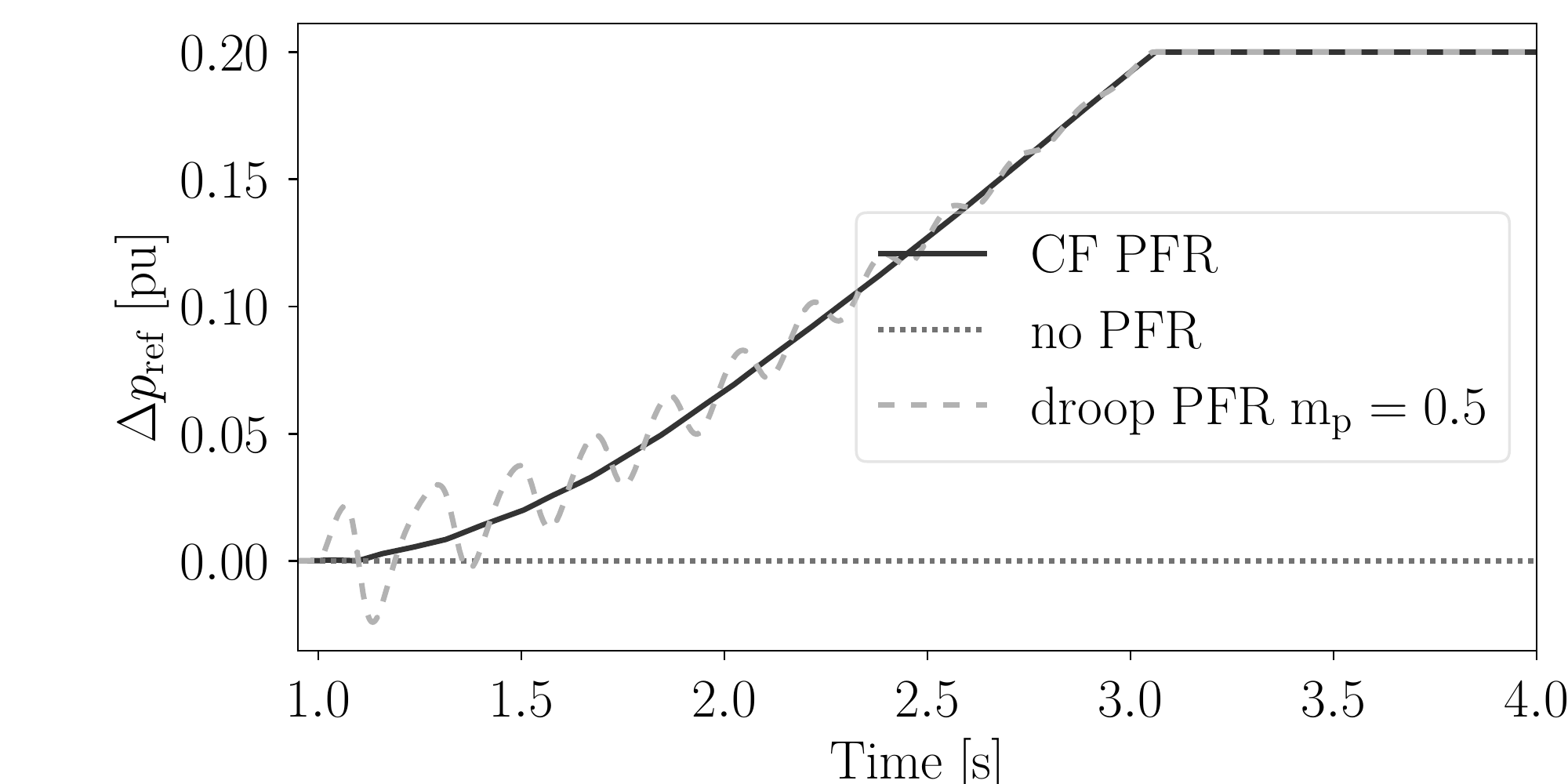}
\includegraphics[width=0.825\columnwidth]{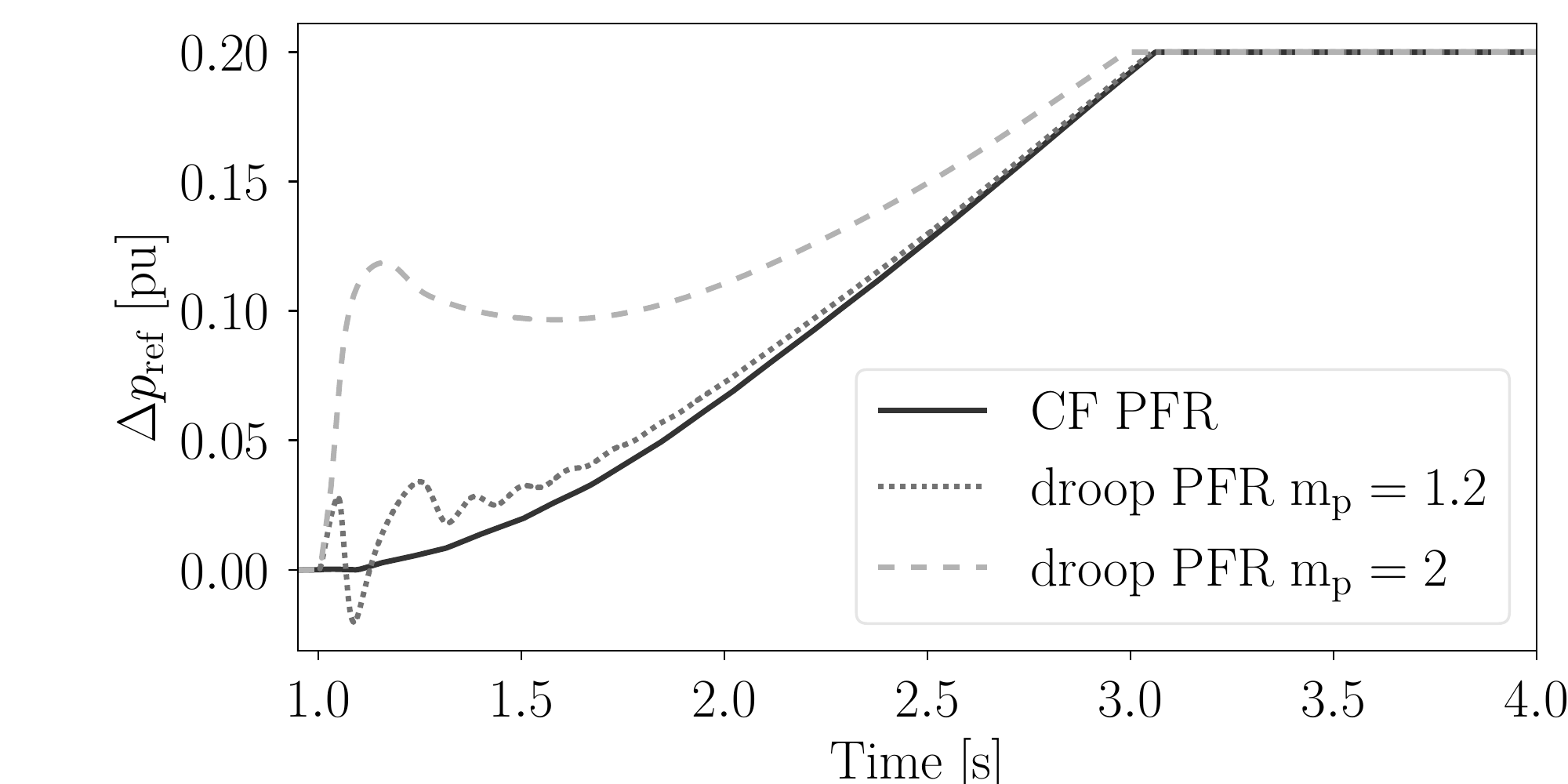}
\caption{Output signal of the frequency control for the \ac{gfm} case. Disconnection at $t=1$~s of the load at bus 5 of the WSCC 9-bus system. The simulation is repeated without frequency control, for two types of input signal and for different values of the active power droop parameter $m_p$.}
\label{fig.pfroutput}
\vspace{-2mm}
\end{figure}

\begin{figure}
\centering
\includegraphics[width=0.825\columnwidth]{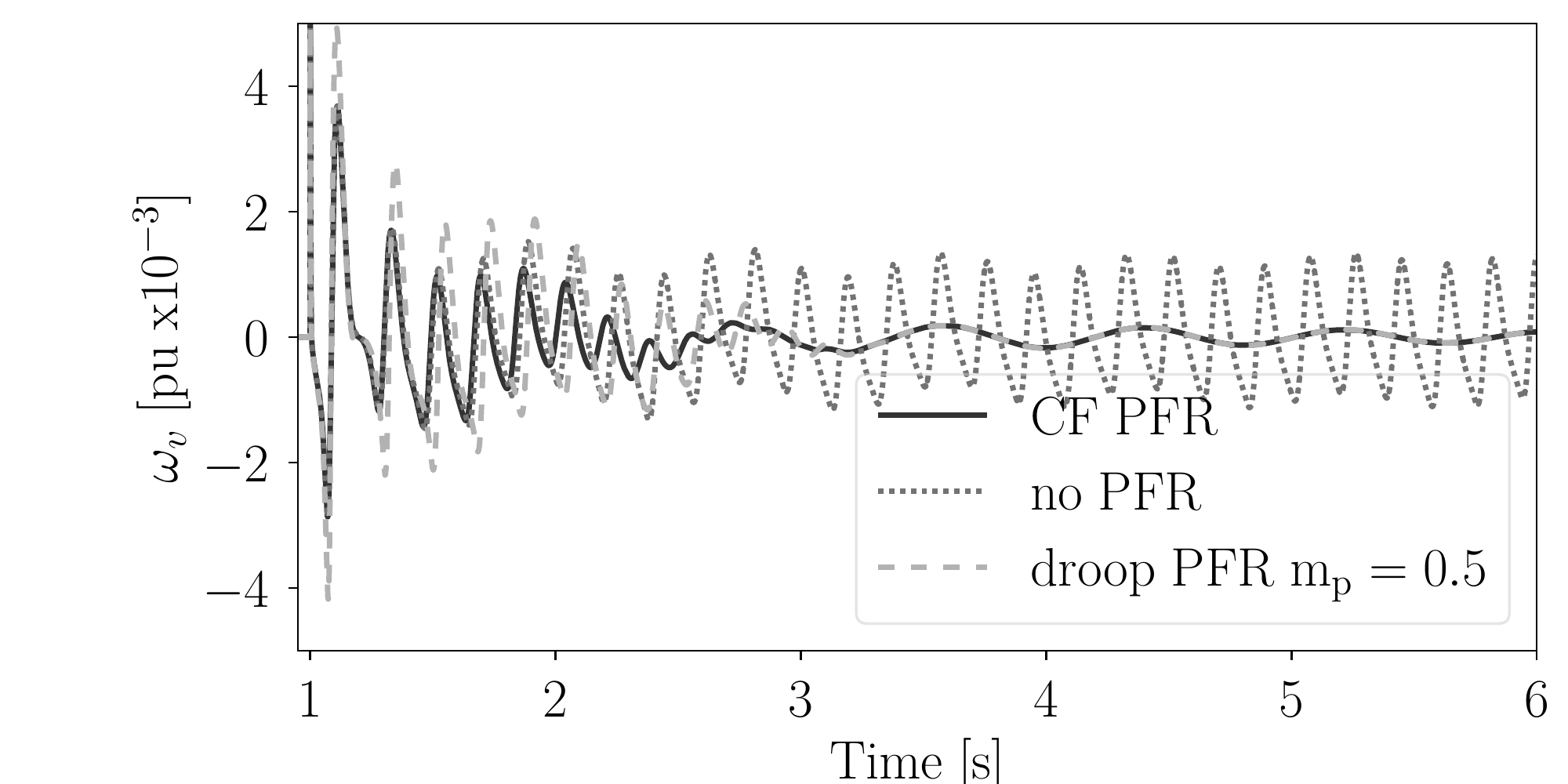}
\includegraphics[width=0.825\columnwidth]{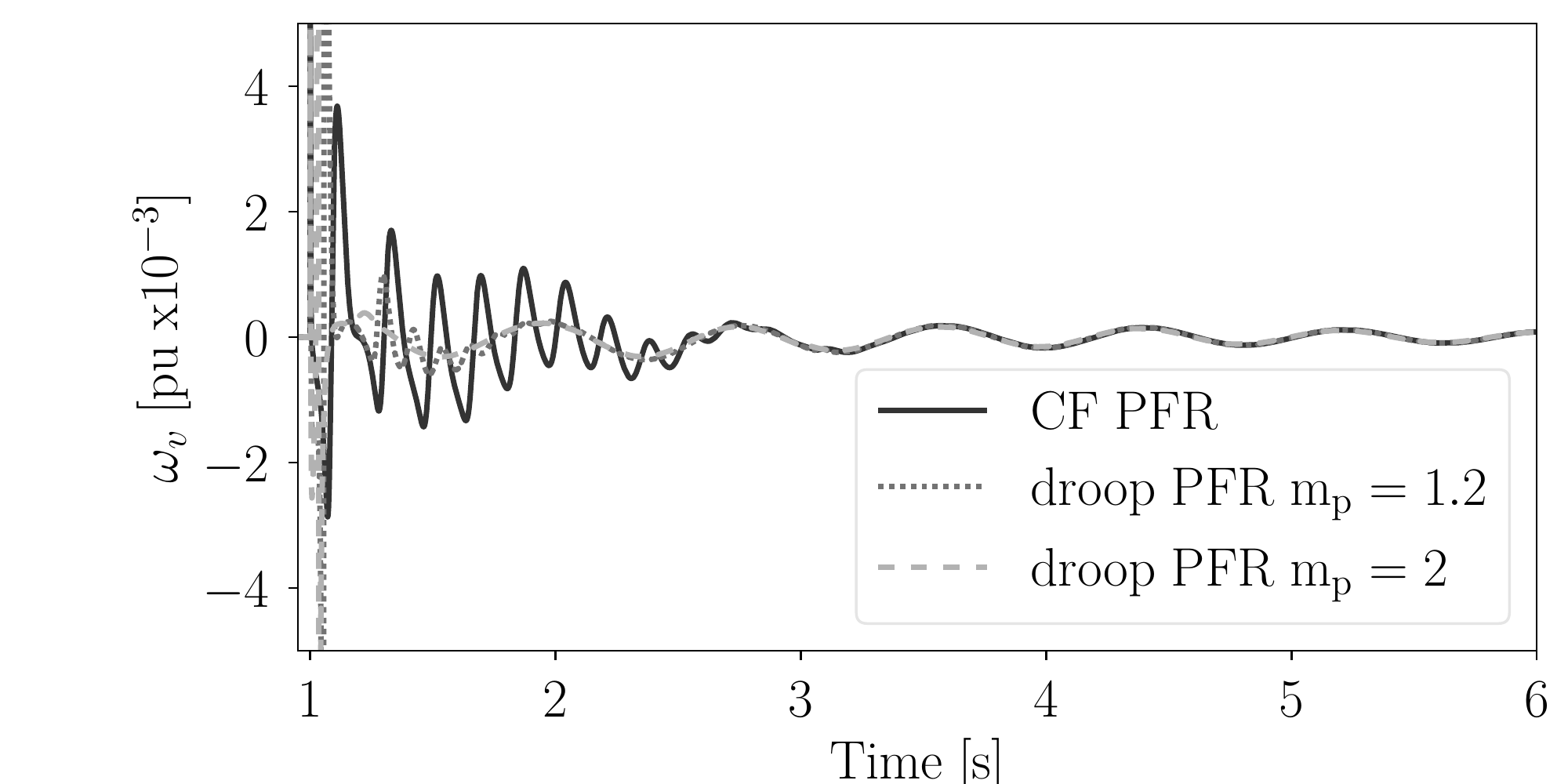}
\caption{Imaginary part of the \ac{cf} as seen at bus 5 of the WSCC 9-bus system during the disconnection at $t=1$~s of the load at bus 5. Application of frequency control for the \ac{gfm} case. The simulation is repeated without frequency control, for two types of input signal and for different values of the active power droop parameter $m_p$.}
\label{fig.pfromegagfm}
\vspace{-2mm}
\end{figure}

Fig.~\ref{fig.pfroutput} shows the output signal of the \ac{pfr} for the \ac{gfm} case.
The exact frequency at the bus is used as well as the internal frequency of the converter with different values of the active power droop gain $m_p$.
It can be seen how the different input signals affects the controller operation until its output is saturated at $t=3$~s.
Fig.~\ref{fig.pfromegagfm} shows the effect of the frequency control at the bus frequency.
Results indicate that, if the droop gain is sufficiently large, the droop \ac{pfr} outperforms the \ac{cf} \ac{pfr} for the first two seconds after the contingency.

\subsection{Comparison between \ac{vsm} and Synchronous Generator}
\label{sub:vsm_sg}
This section illustrates how the \ac{cf} can be used as a
\textit{metric} to directly compare the transient operation of a
\ac{gfm} converter and a synchronous machine.
For this case, the synchronous machine connected to bus 2 of the
network, shown in Fig.~\ref{fig.wscc}, has been substituted with a
\ac{vsm}.
The capacity of the \ac{vsm} and the tuning of its control parameters
are selected in such a way that the \ac{vsm} operates similarly to the
original synchronous machine.  Finally, the same contingency
considered in previous sections is applied also in this scenario.

\begin{figure}[htb]
  \centering
  \includegraphics[width=0.825\columnwidth]{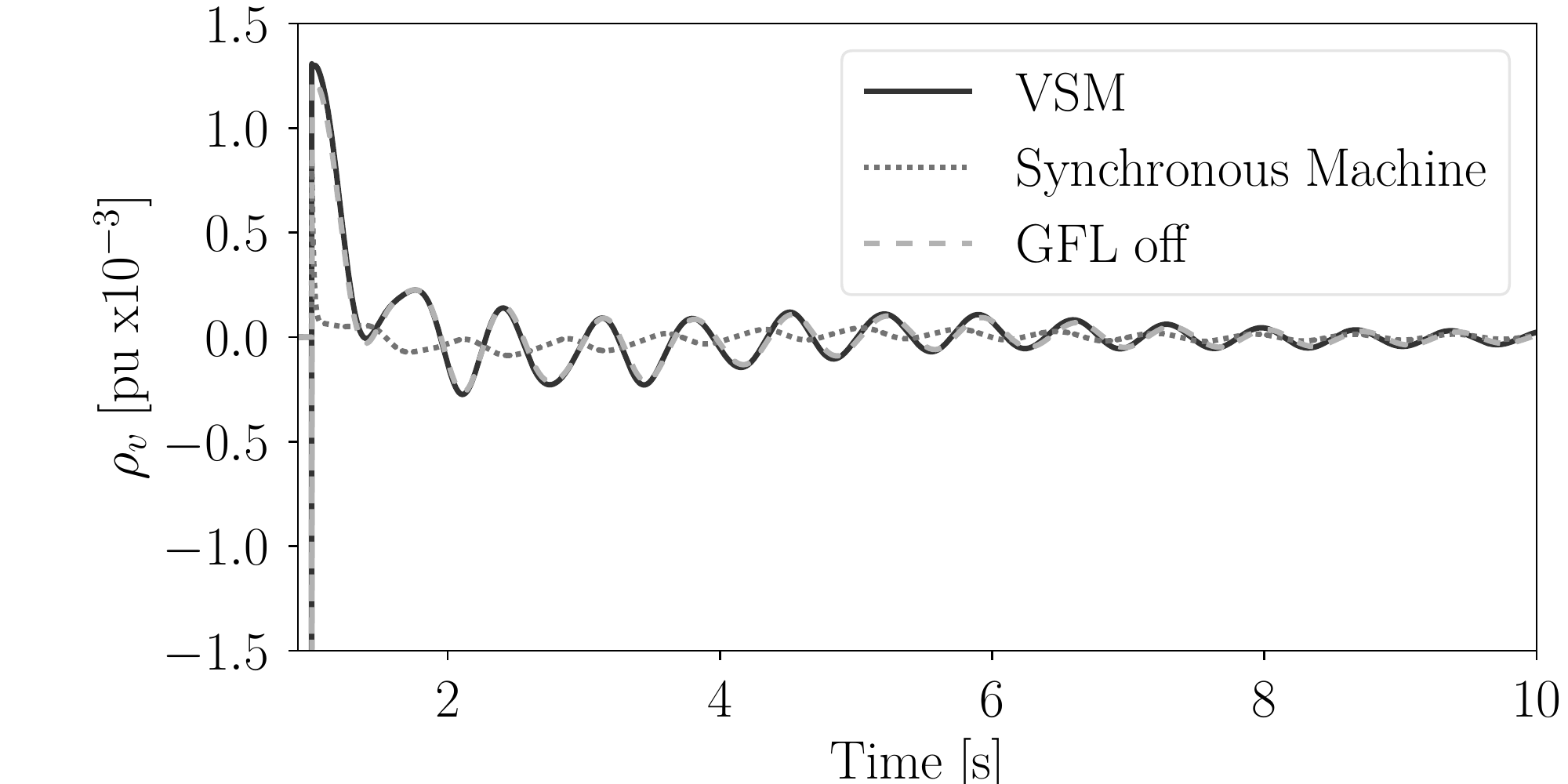}
  \caption{Real part of the \ac{cf} as seen at bus 2 of the WSCC 9-bus
    system during the disconnection at $t=1$~s of the load at bus 5.
    The simulation is repeated with a synchronous generator connected
    to bus 2, a \ac{vsm} substituting the synchronous generator and
    the \ac{gfl} converter at bus 5 disconnected.}
  \label{fig.rhorev}
\end{figure}

\begin{figure}[htb]
  \centering
  \includegraphics[width=0.825\columnwidth]{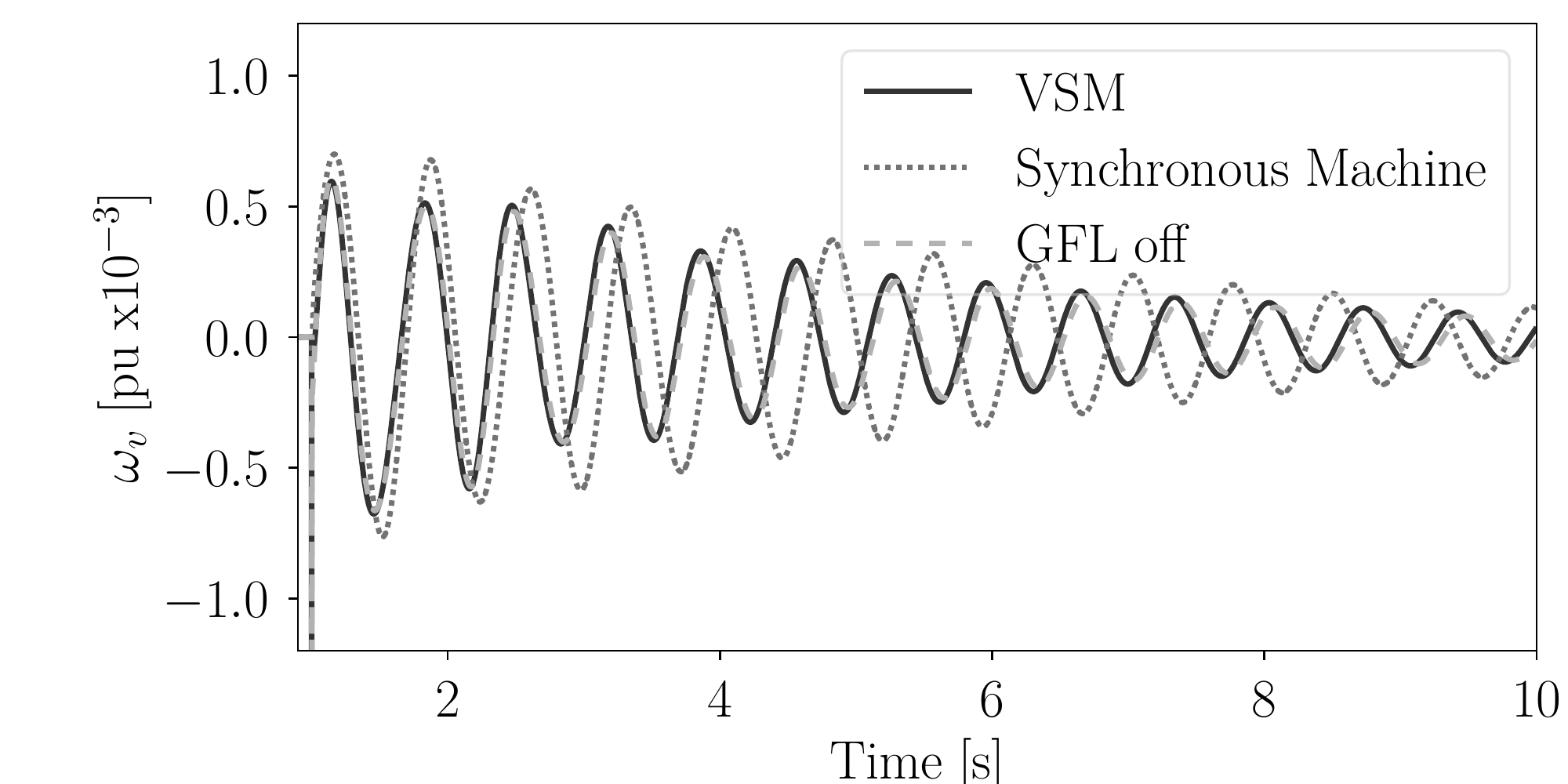}
  \caption{Imaginary part of the \ac{cf} as seen at bus 2 of the WSCC
    9-bus system during the disconnection at $t=1$~s of the load at
    bus 5.  The simulation is repeated with a synchronous generator
    connected to bus 2, a \ac{vsm} substituting the synchronous
    generator and the \ac{gfl} converter at bus 5 disconnected.}
  \label{fig.omegarev}
\end{figure}

Figures~\ref{fig.rhorev} and \ref{fig.omegarev} show the real and
imaginary parts, respectively, of the \ac{cf} at bus 2, in the two
scenarios, namely with synchronous machine and with VSM connected to
the bus.  An additional scenario is shown in these figures, i.e., a
case for which the \ac{gfl} converter at bus 5 is disconnected and the
\ac{vsm} is the only converter in the system. It is shown that the
control and implementation aspects of the \ac{vsm} lead to different
transient operation compared to a traditional synchronous generator.
\ac{cf} captures the discrepancies in the transient operation and can
be used as a metric to directly compare them.
Different trends for the system transient performance are identified
through the use of the different quantities $\rho_v$, $\omega_v$.
Specifically, the magnitude of post-contingency oscillations of
$\rho_v$ is smaller for the synchronous machine while for $\omega_v$,
the use of the \ac{vsm} results in smaller oscillations.
The frequency response with the disconnected \ac{gfl} converter is
nearly identical with the case for which it is connected.
This fact highlights that the interaction between the different
converter control configurations does not affect \ac{cf}
significantly.


\section{Conclusion}
\label{sec.conclusion}
In this paper, the concept of \ac{cf} is utilized to develop a
taxonomy of different power-converter control schemes.
Both \ac{gfl} and \ac{gfm} control configurations are studied and
their effect on the local frequency is analytically derived.
Theoretical results are complemented with a case study based on a
modified model of the WSCC 9-bus system, where the derived analytical
formulations are used for control applications for both \acp{gfl} and
\acp{gfm}.

Results show that \ac{cf} approach decouples the contribution on the
local frequency of each sub-controller and identifies critical control
parameters.
For all converters, the current controller is shown to represent a
constant translation of the real part of the \ac{cf} while the
synchronization control, regardless of its type, affects the imaginary
part.
For \ac{gfl} configurations, the \ac{pll} parameters are shown to have
the largest impact on the local frequency.
For \ac{gfm}, active power droop parameter as well as \ac{vsm} damping
parameter are shown to affect the frequency response after a
contingency.
For the \ac{gfl} control application case, the internal frequency of
the converter, used as an input to a \ac{pfr}, achieves the same
frequency response with the exact frequency measurement, provided that
the \ac{pll} is sufficiently fast.
For the \ac{gfm} case, the internal frequency of the controller
achieves a better transient response than the exact frequency.

In this work, the dynamic effect of conventional controllers on the
frequency at their point of connection was presented and classified.
Future work will focus on extending the use of the calculated internal
frequencies of the converters for control applications.
The potential of using non-conventional controllers based on \ac{cf}
or controllers based on non-conventional input signals, such as the
real part and the magnitude of the \ac{cf}, will also be further
explored.
Finally, the effect on \ac{cf} of multiple converters, their dynamic
interaction and the impact of this interaction on converter frequency
control will also be studied.



\begin{IEEEbiography}[{\includegraphics[width=1in, height=1.25in,
    clip, keepaspectratio]{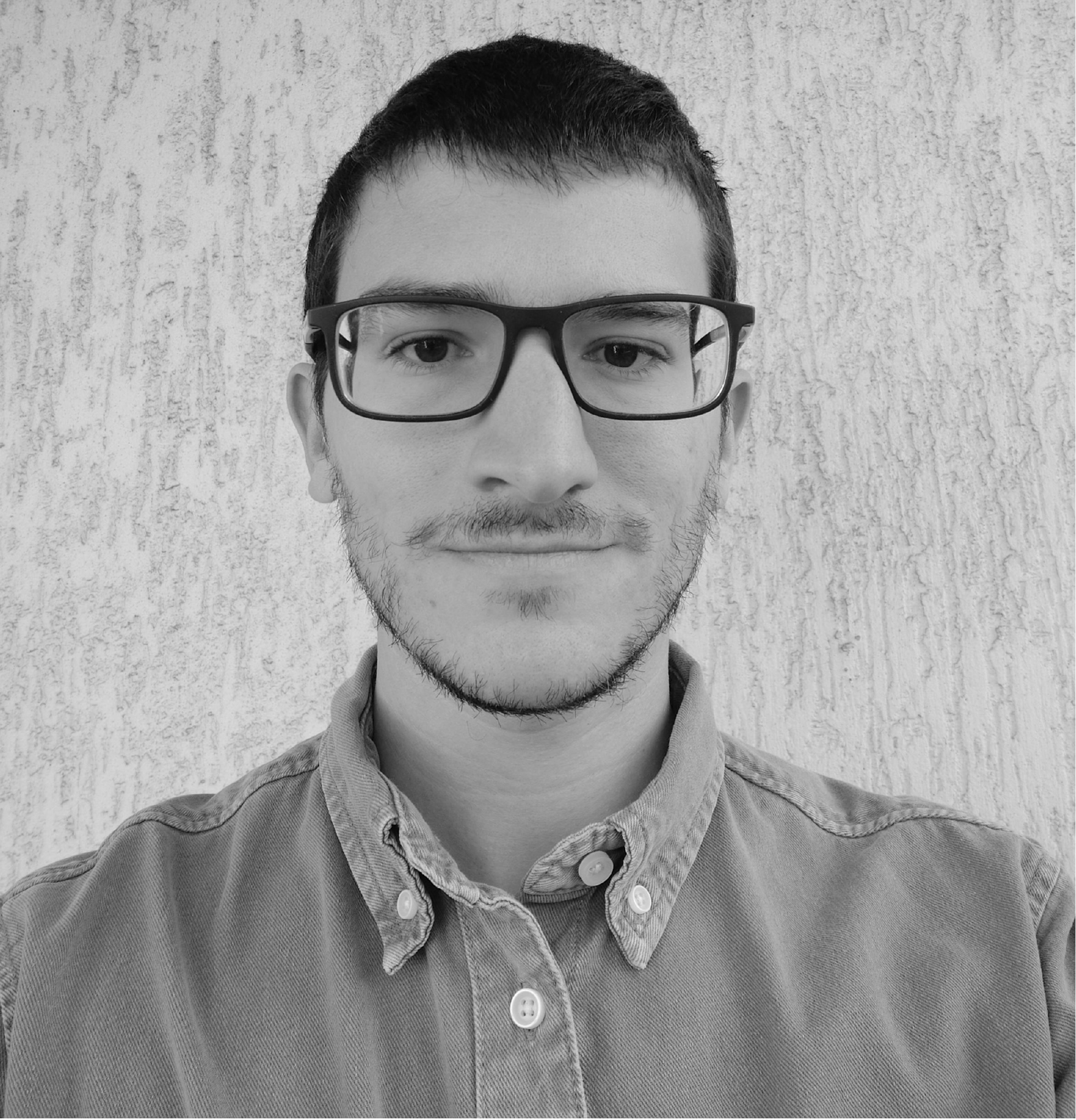}}] {Dionysios
    Moutevelis} received the M.Eng. degree in Electrical and Computer
  Engineering from the National Technical University of Athens, Greece
  in 2017. In 2019 he joined IMDEA Energy Institute, Madrid, Spain
  where he is currently working as a pre-doctoral researcher. From May
  to August 2022 he was with University College Dublin, Ireland, as a
  visiting researcher. His research interests include stability
  analysis of power systems and power converter control.
\end{IEEEbiography}

\begin{IEEEbiography}[{\includegraphics[width=1in, height=1.25in,
    clip, keepaspectratio]{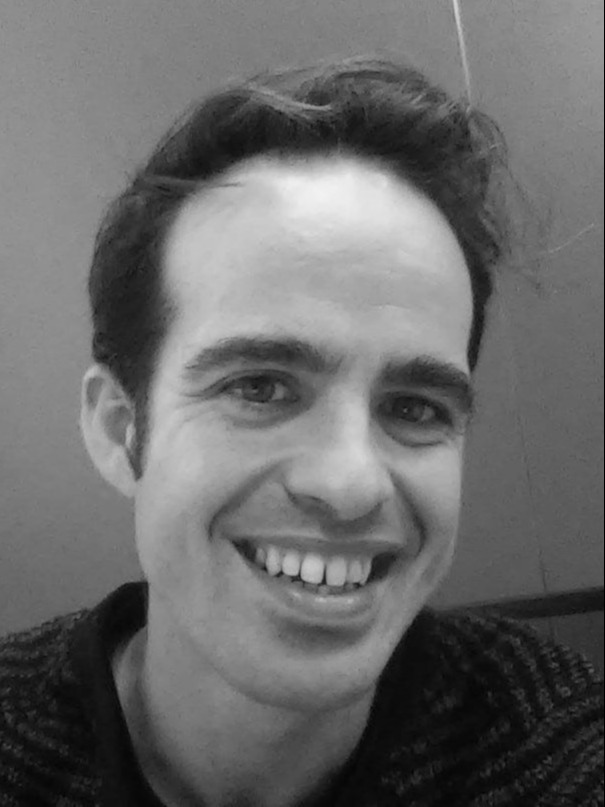}}]%
  {Javier Rold\'{a}n-P\'{e}rez} (S'12-M'14) received a B.S. degree in
  industrial engineering, a M.S. degree in electronics and control
  systems, a M.S. degree in system modeling, and a Ph.D. degree in
  power electronics, all from Comillas Pontifical University, Madrid,
  in 2009, 2010, 2011, and 2015, respectively. From 2010 to 2015, he
  was with the Institute for Research in Technology (IIT), Comillas
  University. In 2014, he was a visiting Ph.D. student at the
  Department of Energy Technology, Aalborg University, Denmark. From
  2015 to 2016 he was with the Electric and Control Systems Department
  at Norvento Energ\'{\i}a Distribuida. In September 2016 he joined
  the Electrical Systems Unit at IMDEA Energy Institute. In 2018, he
  did a research stay at SINTEF Energy Research, Trondheim. His
  research topics are the integration of renewable energies,
  microgrids, and power electronics applications.
\end{IEEEbiography}

\begin{IEEEbiography}[{\includegraphics[width=1in, height=1.25in,
    clip, keepaspectratio]{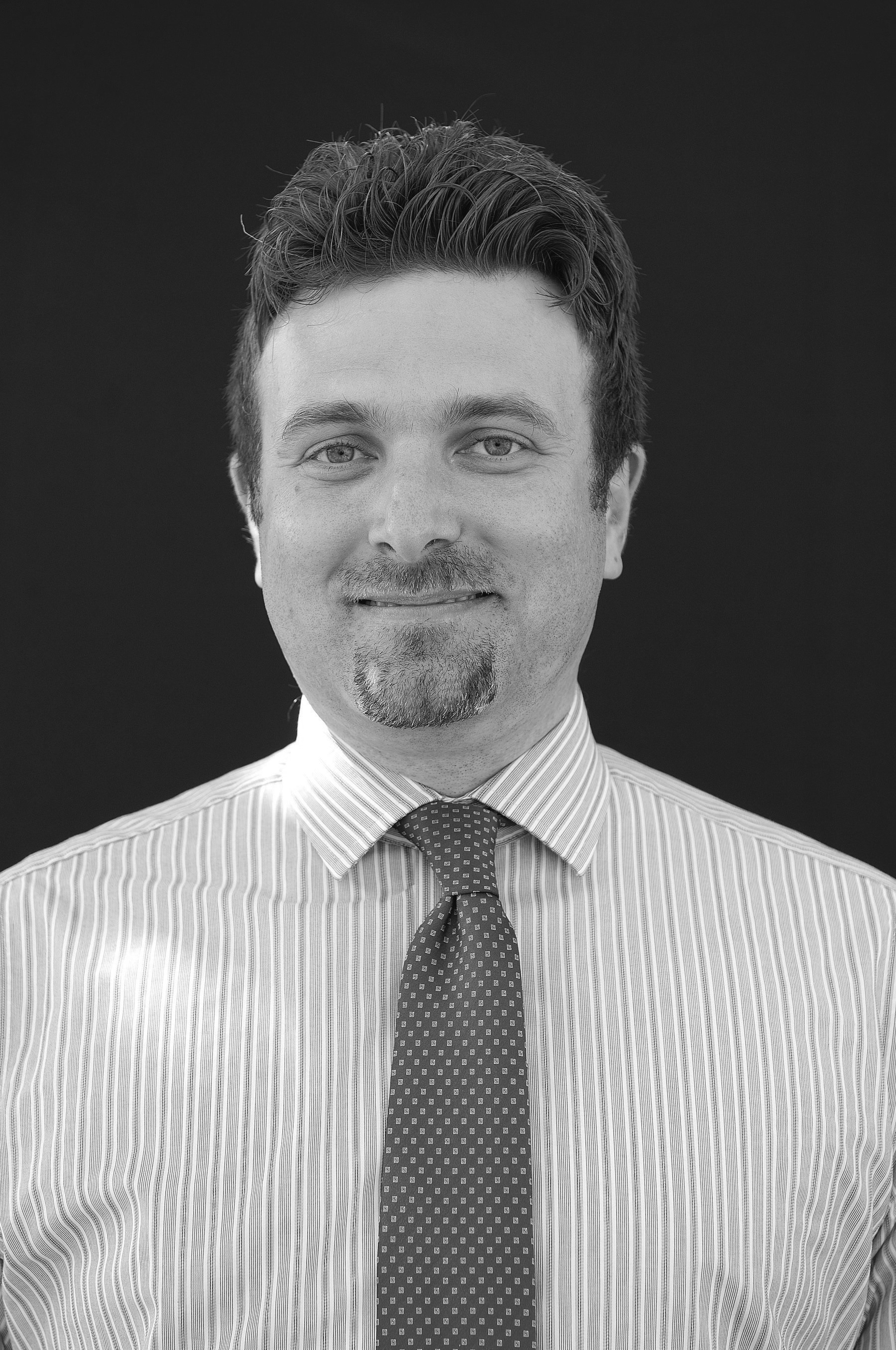}}]%
  {Milan Prodanovic} (Member, IEEE) received the B.Sc. degree in
  electrical engineering from the University of Belgrade, Belgrade,
  Serbia, in 1996 and the Ph.D. degree in electric and electronic
  engineering from Imperial College, London, U.K., in 2004. From 1997
  to 1999, he was with GVS engineering company, Serbia, developing UPS
  systems. From 1999 until 2010, he was a Research Associate in
  electrical and electronic engineering with Imperial College. He is
  currently a Senior Researcher and Head of the Electrical Systems
  Unit, Institute IMDEA Energy, Madrid, Spain. He authored a number of
  highly cited articles and is the holder of three patents. His
  research interests include design and control of power electronics
  interfaces for distributed generation, microgrids stability and
  control, and active management of distribution networks.
\end{IEEEbiography}

\begin{IEEEbiography}
  [{\includegraphics[width=1in, height=1.25in, clip,
    keepaspectratio]{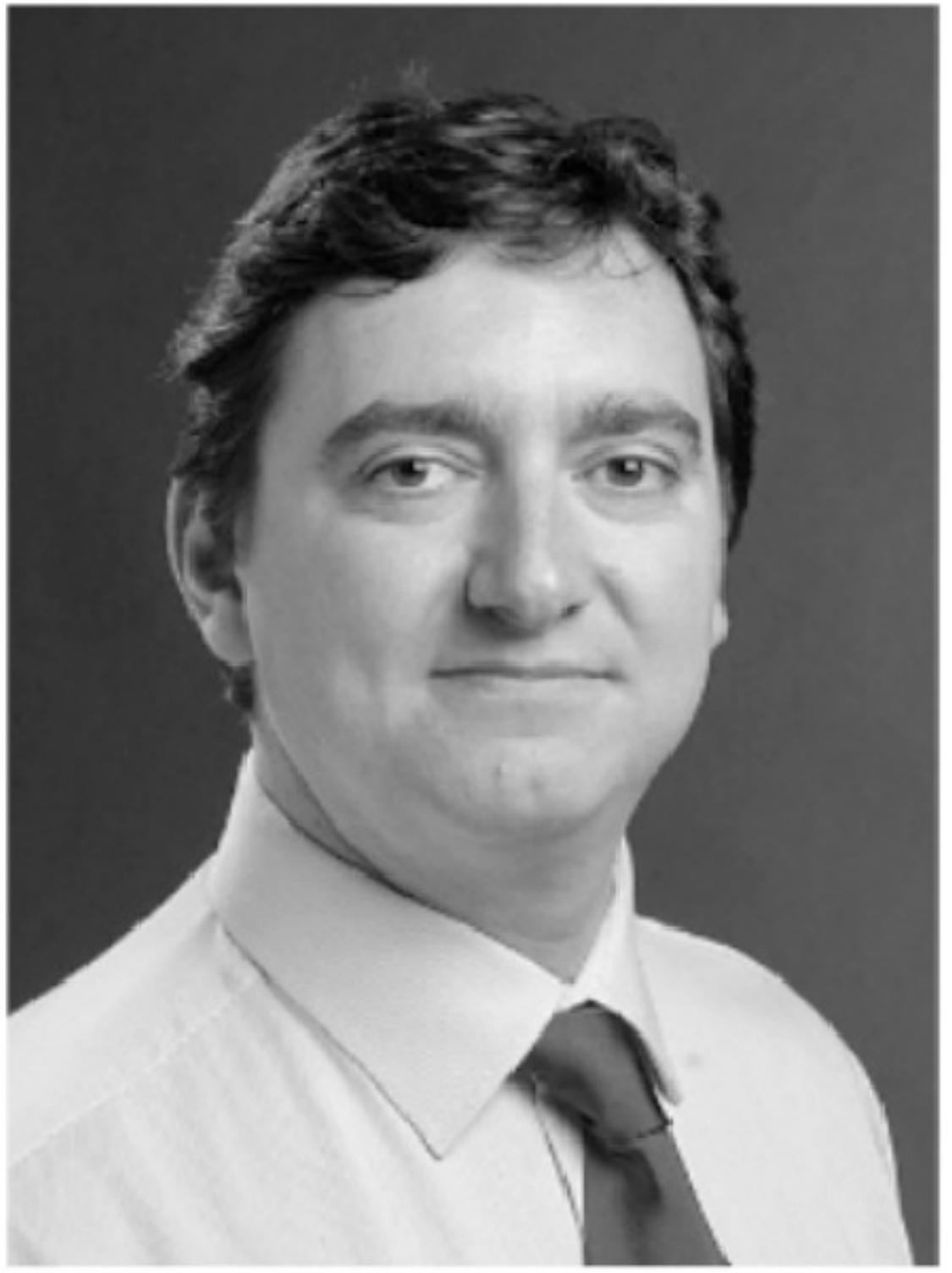}}] {Federico Milano} (F'16)
  received from the Univ. of Genoa, Italy, the ME and Ph.D.~in
  Electrical Engineering in 1999 and 2003, respectively.  From 2001 to
  2002 he was with the University of Waterloo, Canada, as a Visiting
  Scholar.  From 2003 to 2013, he was with the University of
  Castilla-La Mancha, Spain.  In 2013, he joined the University
  College Dublin, Ireland, where he is currently a full professor.  He
  is also Chair of the IEEE Power System Stability Controls
  Subcommittee, IET Fellow, IEEE PES Distinguished Lecturer, Chair of
  the Technical Programme Committee of the PSCC 2024, Senior Editor of
  the IEEE Transactions on Power Systems, Member of the Cigre Irish
  National Committee, and Co-Editor in Chief of the IET Generation,
  Transmission \& Distribution.  His research interests include power
  system modeling, control and stability analysis.
\end{IEEEbiography}

\vfill

\end{document}